\documentclass[twocolumn,12pt]{iopart}
\usepackage{graphicx}
\usepackage{color}
\usepackage{amsbsy}
\usepackage{color}
\usepackage{epsfig}
\usepackage{graphicx}
\usepackage{amssymb}
\usepackage{bbm}
\usepackage{amsbsy}
\usepackage{slashed}
\usepackage{iopams}
\usepackage{xcolor}

\newcommand{\beq}{\begin{equation}}
\newcommand{\eeq}{\end{equation}}

\definecolor{mbcol}{rgb}{1,0,1}

\usepackage[normalem]{ulem}

\begin{document}
\title[]{Color-
flavor dependence of the  Nambu-Jona-Lasinio model and QCD phase diagram}
\author{Aftab Ahmad and Ali Murad}
\address{Institute of Physics, Gomal University, 29220, D.I. Khan, Khyber Pakhtunkhaw, Pakistan.}
\ead{aftabahmad@gu.edu.pk, alimuradeng@gmail.com}

\begin{abstract}
We study the dynamical chiral symmetry breaking/restoration for the various numbers of light
quarks flavors $N_f$ and colors $N_c$, using the Nambu-Jona-Lasinio (NJL) model of quarks, dressed
with a 
color-flavor dependence of effective coupling. Initially, we set $N_f = 2$, 
and varying  the number of colors $N_c$, we find that the dynamical chiral symmetry is broken when $N_c$ exceeds to
its critical value $N^{c}_{c}\approx2.2$. Secondly,  we take  $N_c = 3$, and varying $N_f$, we observed that the dynamical chiral
symmetry is restored when $N_f$ reaches to its critical value $N^{c}_{f}\approx8$. The strong interplay observed
between $N_c$ and $N_f$, i.e.,  $N_c$ anti-screens the strong interactions by strengthening the dynamical
mass and quark-antiquark condensate, while $N_f$ screens the strong interaction by suppressing both
the parameters. We further sketch the quantum chromodynamics (QCD) phase diagram at finite
temperature $T$ and quark chemical potential  $\mu$ for various $N_c$ and $N_f$. At finite $T$ and  $\mu$, we
observed that the critical number of colors $N^{c}_c$ enhances while the critical number of flavors $N^{c}_f$
suppresses as  $T$ and $\mu$ increases. Of course,  the parameters $T$ and $\mu$ produces the screening effect. Consequently, the critical temperature $T_c$, $\mu_c$ and
co-ordinates of the critical endpoint $(T^{E}_c,\mu^{E}_c)$ in the QCD phase diagram enhances as $N_c$ increases
while  suppresses when $N_f$ increases. Our findings agree with the Lattice QCD  and  Schwinger-Dyson equations predictions.
\end{abstract}


%\textit {Published in Chin. Phys. C 45,5}

\maketitle

\section{Introduction}\label{section-I}
Quantum Chromodynamics (QCD) is a well-established theory of strong color interaction
among the quarks and gluons. Two major aspects
of the QCD are the asymptotic freedom (ultraviolet
regime)~\cite{Gross:1973id, Politzer:1973fx} and quark confinement(infrared regime)~\cite{Wilson:1974sk}.
In the asymptotic freedom, the quarks interact weakly at
a short distance inside the hadrons. On the other hand,
at a large distance (or at low energy)  the quarks are
confined and never exist in isolation. Besides the color
confinement, the dynamical chiral symmetry breaking
is another important property of the low-energy QCD,
which is related to the dynamical mass generation of the
quarks.
It is well known that in the fundamental representation
of the gauge group $SU(N_c)$, the QCD exhibits confinement and dynamical chiral symmetry breaking with 
the small number of light quark 
flavors $N_f$. But for larger
$N_f$
, it is believed that there is a critical value
$N^{c}_{f}$
above which, the chiral symmetry restores, and quarks become
unconfined~\cite{Appelquist:2009ty, bashir2013qcd, LSD:2014nmn}. This  value $N^{c}_{f}$
, must be smaller
than the upper limit of the critical value where asymptotic freedom is appeared to exists is $N^{A,c}_{f}= (11/2)N_c$ ~\cite{Politzer:1973fx}, for $N_c=3$ with a gauge group $SU(3)$, this critical number is $N^{A,c}_{f}=16.5$. Hence, the 
QCD theory is said to be conformal in the infrared,
guided by an infrared fixed point (i.e., a point at which
the $\beta$-functions for the QCD couplings vanishes),
see for 
example in detail~\cite{Caswell:1974gg,Banks:1981nn,gies2006chiral,Appelquist:2007hu,hasenfratz2010conformal,aoki2012many} . The region $N^{c}_{f}\lesssim N_f<N^{A,c}_{f}$, is often called as the ``conformal zone''~\cite{Appelquist:2009ty, LSD:2009yru}. At or near the upper end $(N_f \lesssim N^{A,c}_{f})$
of the conformal zone, the infrared fixed point lies in
the weakly interacting region and can be dealt with
perturbative techniques of QCD. On the other hand,
around the lower end $(N_f \sim N^{c}_{f})$, the infrared fixed
point shift toward the strongly interacting region where
the coupling is sufficiently strong as $N_f$ decreases and
thus, the system enters a phase where the chiral
symmetry breaks and quarks become confined. 
In this situation, the perturbative approaches to QCD are
inconceivable and thus, the non-perturbative techniques
are  useful tool. Lattice QCD simulations~\cite{LSD:2014nmn,Hayakawa:2010yn,Cheng:2013eu,Hasenfratz:2016dou,LatticeStrongDynamics:2018hun},
as well as the continuum methods of QCD ~\cite{bashir2013qcd,Appelquist:1999hr,Hopfer:2014zna,Doff:2016jzk,Binosi:2016xxu,Ahmad:2020jzn}, in the fundamental $SU(3)$ representation, emphasized that the chiral symmetry restoration and deconfinement phases exiting when $N_f$ reaches to the conformal zone $8\lesssim N^{c}_f< 12$.
QCD theory with a larger number of colors $N_c$, in the
fundamental $SU(N_c)$,  representation, also plays a significant 
role in the infrared domain.  It has
been discussed in ~\cite{Ahmad:2020jzn}, that the dynamical chiral
symmetry also  breaks when $N_c$ exceeded
to a critical value $N^{c}_{c}\approx2.2$.
The increasing number
of colors $N_c$, anti-screens the strong interactions and
enhances the dynamical mass generation near and above $N^{c}_{c}$.
The large $N_{c}$, also effects the  critical value of flavors $N^{c}_{f}$
, that is, $N^{c}_{f}$
shifts
toward its higher values as $N_{c}$ increases. The $N_{c}$ and
$N_{f}$ compete each other, i.e., $N_f$ screens the  strong interactions in contrast to the  $N_c$ anti-screening effect~\cite{Ahmad:2020jzn}.\\
Besides the $N_{c}$ and $N_{f}$
, the infrared domain of QCD is
also affected by the presence of heat bath $T$. It is believed that at zero or low $T$, the fundamental degrees of
freedom of low energy QCD are the color-singlet (confined) hadrons, whereas, at high $T$, when $T$ reaches to its critical value
$T_c$, the interaction becomes weak and thus, causing hadrons to melt down into a new
phase, where now the quarks and gluons becomes the new degrees of freedom. The chiral symmetry
is restored and deconfinement of quarks occurs in this new phase. Lattice QCD calculation~\cite{Bernard:2004je,Cheng:2006qk,Bazavov:2011nk,Bhattacharya:2014ara,deForcrand:2014tha,Bazavov:2017xul,Guenther:2020jwe}, Schwinger-Dyson equation~\cite{Qin:2010nq,Fischer:2011mz,Ayala:2011vs,gutierrez2014qcd,Eichmann:2015kfa,Gao:2016qkh,ahmad2016inverse,Fischer:2018sdj,Shi:2020uyb,Ahmad:2020ifp} and another effective model of
QCD~\cite{klevansky1992nambu,buballa2005njl,costa2010phase,marquez2015dual,Ahmad:2015cgh,Ayala:2017gek,Ayala:2021nhx} claims that the nature of this transition is
cross-over in the presence of finite current quark mass
$m$. Similarly, upon increasing quark chemical potential
$\mu$, the same physical picture prevails, but the nature of
the phase transition changes from cross-over to the first-order at  some point, known as the critical endpoint in the
QCD phase diagram, usually drawn in the $T-\mu$ plane.
Experimentally, there is a strong motivation regarding
the study of phase transition in the Heavy-Ion Collision
at the Large Hadron Collider (LHC) in
CERN, the Relativistic Heavy Ion Collider (RHIC) at
Brook Heaven National Laboratory (BNL), and Compact Baryonic Matter (CBM) experiments.  The study of low energy QCD for higher quark  flavors $N_f$ and $N_c$ also plays 
a significant role in physics beyond the standard
model (BSM), and its extension to the QCD phase diagram
at finite $T$ and $\mu$,  where different phases of QCD exists,
i.e., quark-gluon plasma, quarkonia matter, neutron
star environment and the color-favor-locked region of
the QCD phase diagram.\\
Our aim and motivation of this work  is  to investigate
the critical number of colors $N_c$ and flavors $N_f$ for the chiral symmetry breaking and restoration at zero $T$, finite $T$ and $\mu$.  Furthermore, we are interested to draw   QCD phase diagram in the $T-\mu$ plane for various $N_c$ and $N_f$.  We use the effective Nambu-Jona-Lasinio (NJL) model~\cite{nambu1961dynamical}, in the Hartree-Fock mean-field approximation, dressed with the number of quark flavors $N_f$ and colors $N_c$. NJL model has
several features and is commonly used in the
literature. The Dynamical chiral symmetry breaking and its
restoration are some of the main features of this model.
However, it does not support the phenomenon of quark
confinement.\\
This article is organized as follows: In Sec.~\ref{section-II}, we
introduce the general formalism of the NJL model at
$T=0$, at  finite $T$ and  $\mu$.
We discuss the numerical solution of the gap equation to study the dynamical
chiral symmetry breaking and its restoration for higher 
colors $N_c$ and  flavors $N_f$  in Sec.~\ref{section-III}. We present the numerical solution of the gap equation for higher $N_c$ and $N_f$ and at finite temperature $T$ in Sec.~\ref{section-IV}. In Sec.~\ref{section-V}, we  display the  numerical solution of  the gap equation at finite chemical potential $\mu$, for  various  $ N_c$  and $N_f$.  We sketch the phase diagram in the $T-\mu$ plane for various $N_c$ and $N_f$ in the Sec.~\ref{section-VI}.
In the last Sec.~\ref{section-VII}, we demonstrate the summary and
conclusions of this work.
\section{General Formalism of the  NJL Model}\label{section-II}
We start QCD  in an effective manner through NJL Lagrangian~\cite{nambu1961dynamical}:
\begin{equation}
{\cal L}=\bar{q} (i\not \! {\partial}-m)q + \frac{G}{2}[(\bar{q}q)^2+(\bar{q} i\gamma_5 \vec{\tau}q)^2],\label{NJL}
\end{equation}
where the four-Fermi interactions term contains a scalar and an axial-vector interaction piece ($\tau$ representing the Pauli matrices in isospin space) and $G$ is the coupling of the theory. Such a Lagrangian~Eq.~(\ref{NJL}) describes the dynamical chiral symmetry breaking, which can be triggered through Schwinger-Dyson equations (SDE) for the dressed quark propagator, can be written as  
\begin{eqnarray}
S^{-1}(p)&=& S^{-1}_{0}(p) + \Sigma(p). \label{NJL1}
\end{eqnarray}
Here $S^{-1}$ is the inverse of the  dressed quark propagator:
\begin{eqnarray}
S^{-1}(p)=i\gamma \cdot p + M, \label{NJL2}
\end{eqnarray}
with $M$ is the dynamical quark mass, $\gamma^{\mu}$ are the Dirac $4\times4$ gamma matrices and $p^{\mu}$ are the four-momenta. The inverse of  the free quark propagator  $S^{-1}_{0}$ is defined as
\begin{eqnarray}
S^{-1}_{0}(p)=i\gamma \cdot p + m, \label{NJL3}
\end{eqnarray}
where $m$ is the bare quark mass, which we may set equal to zero in the chiral limit.  The $\Sigma(p)$ is the quark  self energy which can be written as 
\begin{eqnarray}
\Sigma(p)=
\int \frac{d^4k}{(2\pi)^4} g^2
 \Delta_{\mu\nu}(p-k)\frac{\lambda^a}{2}\gamma_\mu S_f(k)
\frac{\lambda^a}{2}\Gamma_\nu(p,k)\,,\label{NJL4}
\label{eqn:gap-QCD}
\end{eqnarray}
here $g$ is the QCD coupling constant, $\Gamma_\nu $  is the dressed quark-gluon vertex,
$\Delta_{\mu\nu}$ is the gluon propagator and $S(k)$ is the dressed quark propagator:
\begin{equation}
S(k) = \frac{i\gamma \cdot p + MI}{p^2 - M^2 + i\epsilon},
\label{NJL5}
\end{equation}
here $ i\epsilon$ is the causality  factor, introduced to exclude the singularity from the propagator,
the $\lambda^{a}$ are the
usual Gell-Mann's matrices. In the $SU(N_c)$ representation the Gell-Mann's  matrices satisfies the following identity:
\begin{eqnarray}
\sum^{8}_{a=1}\frac{{\lambda}^a}{2}\frac{{\lambda}^a}{2}=\frac{1}{2}\left(N_c - \frac{1}{N_c} \right). \label{NJL6}
\end{eqnarray} 
In NJL model,  we set
\begin{eqnarray}
  g^2 \Delta_{\mu \nu}&=& G\delta_{\mu \nu},\label{NJL7}
\end{eqnarray}
%The mathematical expression corresponding to the Fig. (\ref{DSEs}) is given by Eq. (\ref{2.11}),
where $G$ is the effective coupling. The effective coupling $G$  must exceeds a critical value $G_c$, in order to describes the  dynamical chiral symmetry breaking. When $G$ is greater then its critical value $G_c$, a nontrivial solution to the  QCD gap equation bifurcates from the trivial one, see for example Ref.~\cite{Ahmad:2018grh}. 
Substituting Eqs.~{(\ref{NJL2}),(\ref{NJL3}),(\ref{NJL4}),(\ref{NJL5}),(\ref{NJL6}),(\ref{NJL7})}, in Eq.~(\ref{NJL1}) and taking the trace over  the Dirac, colors and flavors  components, the quark gap equation is given as 
\begin{eqnarray}
M = m + 8i \mathcal{G}^{N_c}(N_f)\int{\frac{d^{4}k}{(2\pi)^4}\frac{M}{k^2 - M^2 + i\epsilon}},
\label{NJL8}
\end{eqnarray}
%\begin{equation}
%M=m_f+2i \mathcal{G}^{N_c}(N_f)\int \frac{d^4k}{(2\pi)^4} S(p)\;,\label{NJL7}
%\end{equation}
where $M$ is the dynamical mass  and $\mathcal{G}^{N_c}(N_f) $ is the  effective coupling  in which the  color and flavor factors are incorporated as
\begin{eqnarray}
\mathcal{G}^{N_c}(N_f)=[\frac{1}{2}\left(N_c -\frac{1}{N_c}\right)]G(N_f).\label{NJL9} 
\end{eqnarray} 
To study the gap equation for the various number of 
flavors, we needs to modify the flavor sector $G(N_f)$ of the effective coupling $\mathcal{G}^{N_c}(N_f)$, in such a way that it must give solutions of the gap equation for the higher number of flavors. We shall use this modification as our modeling and it will help us to find out the critical number of flavors $N^{c}_{f}$ or colors $N^{c}_{c}$ for the chiral symmetry breaking or restoration. We adopted a similar way  of modeling used by~\cite{bashir2013qcd, Ahmad:2020jzn}, where the critical number of flavors $N^{c}_{f}$
obtained from Schwinger-Dyson equations. According to Ref.~\cite{bashir2013qcd, Ahmad:2020jzn}, the dynamically generated
mass $M$ should have the following kind of relationship
with the critical number of flavors $N^{c}_{f}$:
\begin{eqnarray}
M \backsim \sqrt{1 - \frac{N_f}{N^{c}_{f}}}, \label{NJL10}
\end{eqnarray}
where $N^{c}_{f}$ denote the critical number of flavors. To find the critical number of flavors $N^{c}_{f}$, we modified the factor 
$G(N_f)$ of Eq.~(\ref{NJL9}) in  similar fashion 
as ~\cite{Ahmad:2020jzn}:
\begin{eqnarray}
G(N_f) \longrightarrow \frac{9}{2}G\sqrt{1 - \frac{(N_{f}-2)}{\mathcal{N}_{f}^{c}}}, 
\label{NJL11}
\end{eqnarray}
with $\mathcal{N}_{f}^{c}=N^c_{f}+\eta$, is some guess values of critical number of 
flavors.  Our modification of the effective coupling
is almost the same as in Ref.~\cite{Ahmad:2020jzn}, but slightly different by a factor of $9/2$, this is because, the effective coupling model of Ref.~\cite{Ahmad:2020jzn},  uses a  symmetry preserving, local four-point contact interaction  model of quarks with the quantum numbers of a massive-gluon
exchange and in the scalar-pseudoscalar channel this is
Fierz equivalent to the NJL-Lagrangian Eq.~(\ref{NJL})~\cite{marquez2015dual}.\\
To obtain the $N^{c}_{f}$, we set $\eta=2.3$, which lies in the range
as predicted in~\cite{Ahmad:2020jzn}  by considering $N^{c}_{f}=8$. It has been
demonstrated in the Ref.~\cite{Ahmad:2020jzn} that the appearance of the parameter $\eta$ is
because of the factor $(N_f-2)$ in Eq.~(\ref{NJL11}).  For fixed
$N_c =3$ and $N_f =2$, our modified NJL effective coupling
$\mathcal{G}^{N_c}(N_f)\rightarrow GN_f N_c$, normally used in NJL model gap
equation~\cite{buballa2005njl}. \\
The four-momentum integral in Eq.~(\ref{NJL8}), can
be solved by splitting the four-momentum into time and
space components. We denote the space part by a bold
face latter $\textbf{k}$ and the time  part by $k_0$. Thus, the Eq.~(\ref{NJL8}), can be written as
\begin{eqnarray}
M = m + 8i \mathcal{G}^{N_c}(N_f) M\int_{0} ^{\infty}\frac{d^3 \textbf{k}}{(2\pi)^4} \int_{-\infty} ^{+\infty} {\frac{dk_{0}}{ {k_{0}} ^{2} - E_{k} ^{2} + i\epsilon}}.
\label{NJL12}
\end{eqnarray}
Here, $E_{k}=\sqrt{{|\textbf{k}}|^2 + M^2}$, in which $E_{k}$ denotes the energy per particle and $\textbf{k}$ is the $3$-momentum. On integrating over the time component of Eq.~(\ref{NJL12}), we can get the following expression:
\begin{eqnarray}
M = m + 8i \mathcal{G}^{N_c}(N_f) M \int_{0} ^{\infty} \frac{d^3 \textbf{k}}{(2\pi)^4} \frac{\pi}{i E_{k}}
\label{NJL113}
\end{eqnarray}
In spherical polar coordinates $d^3 \textbf{k} = \textbf{k}^2 d\textbf{k} sin\theta d\theta d\phi$ and performing the angular integration,  we have  from Eq.~(\ref{NJL113}):
\begin{eqnarray}
M =  m + \frac{2 \mathcal{G}^{N_c}(N_f) M}{\pi^2} \int_{0} ^{\infty}d \textbf{k} \frac{\textbf{k}^2}{E_{k}}.
\label{NJL14}
\end{eqnarray}
The integral occurring in Eq.~(\ref{NJL11}) is a diverging integral and we also know that the NJL model is not renormalizable due to fermionic contact interaction.  One can find different kinds of regularization schemes used in the literature \cite{klevansky1992nambu}.  The regularization procedure we adopted in the
present scenario is the three-dimensional (3d)  momentum cut-off,  where we remove the divergence by applying a certain high ultraviolet 3d-momentum cut-off  $\Lambda $.  So,  Eq.~(\ref{NJL14}) can be written as
\begin{eqnarray}
M =  m + \frac{2\mathcal{G}^{N_c}(N_f) M}{\pi^2} \int_{0} ^{\Lambda}d \textbf{k} \frac{\textbf{k}^2}{E_{k}}.
\label{NJL15}
\end{eqnarray}
On integrating Eq.~(\ref{NJL15}), we have
\begin{eqnarray}
 M =  m +\frac{\mathcal{G}^{N_c}(N_f) M}{\pi^2} \left[\Lambda\sqrt{{\Lambda}^2 + M^2} - M^2 {\rm arcsinh}(\frac{\Lambda}{M})\right].
\label{NJL16}
\end{eqnarray}
In the present situation, the quark-antiquark condensate which serve as an order parameter for the chiral symmetry
breaking is defined as
\begin{eqnarray}
-\langle \bar{q} q\rangle= \frac{M-m}{2\mathcal{G}^{N_c}(N_f)}.\label{NJL17}
\end{eqnarray}
\\
The finite temperature $T$ and quark chemical potential $\mu$ version of the  NJL-model gap equation Eq.~(\ref{NJL8}), can be obtained by adopting the standard convention for momentum integration i.e.,
\begin{eqnarray}
\int\frac{d^4k}{i(2\pi)^4} f(k_0,{\bf{k}})\rightarrow T \sum_{n} \int\frac{d^3 k}{(2\pi)^3}f(i\omega_n+\mu,\bf{k}), \label{NJL18}
\end{eqnarray}
with $\omega_n = (2n+1)\pi T$ are  the fermionic Matsubara frequencies. After doing some algebra, 
the gap equation Eq.~(\ref{NJL8}), at  finite
 $T$ and $\mu$, can be written as
\begin{eqnarray}
M = m + 4\mathcal{G}^{N_c}(N_f)M\int^{\Lambda}_{0} \frac{d^{3}\textbf{k}}{(2\pi)^{3}}\frac{1}{E_{k}} (1 - n_{F}(T, \mu)- \bar{n}_{F} (T, \mu)),
\label{NJLT19}
\end{eqnarray}
which is similar to the Eq.~(\ref{NJL16}) in vacuum, but modified by the thermo-chemical parts. 
Where $ n_F(T, \mu) $ and $ \bar{n}_F(T, \mu) $ represents the Fermi occupation numbers for the quarks and antiquark, respectively, and defined as
\begin{eqnarray}
n_{F}(T, \mu) = \frac{1}{e^{(E_{k}-\mu)/T} + 1} \hspace{3mm}, \hspace{3mm} \bar{n}_{F}(T, \mu)= \frac{1}{e^{(E_{k}+\mu)/T} + 1}.
\label{NJL20}
\end{eqnarray}
On further simplifying, we have
\begin{eqnarray}
M =&  m + \frac{\mathcal{G}^{N_c}(N_f) M}{\pi^2} \left[\Lambda\sqrt{{\Lambda}^2 + M^2} - M^2 {\rm arcsinh}(\frac{\Lambda}{M})\right]\nonumber\\- & \frac{\mathcal{G}^{N_c}(N_f)M}{\pi^{2}}\int_{0}^{\Lambda} d\textbf{k} \frac{\textbf{k}^2}{E_{k}} \left[ n_{F}(T, \mu)+ \bar{n}_{F} (T, \mu)\right]
\label{NJL21}
\end{eqnarray}
If we set $ \mu = T = 0 $ in Eq.~(\ref{NJL21}), we get $n_{F}=\bar{n}_{F}=0$  
and thus, we can retain the gap equation in vacuum i.e., Eq.~(\ref{NJL16}).
In the next section, we present numerical solution of the  NJL-model gap equation for higher number of colors and flavors .

\section{Dynamical chiral symmetry breaking/restoration for $N_c$ and $N_f$}\label{section-III}
\begin{figure}[h!]
\begin{center}
\includegraphics[scale=0.4]{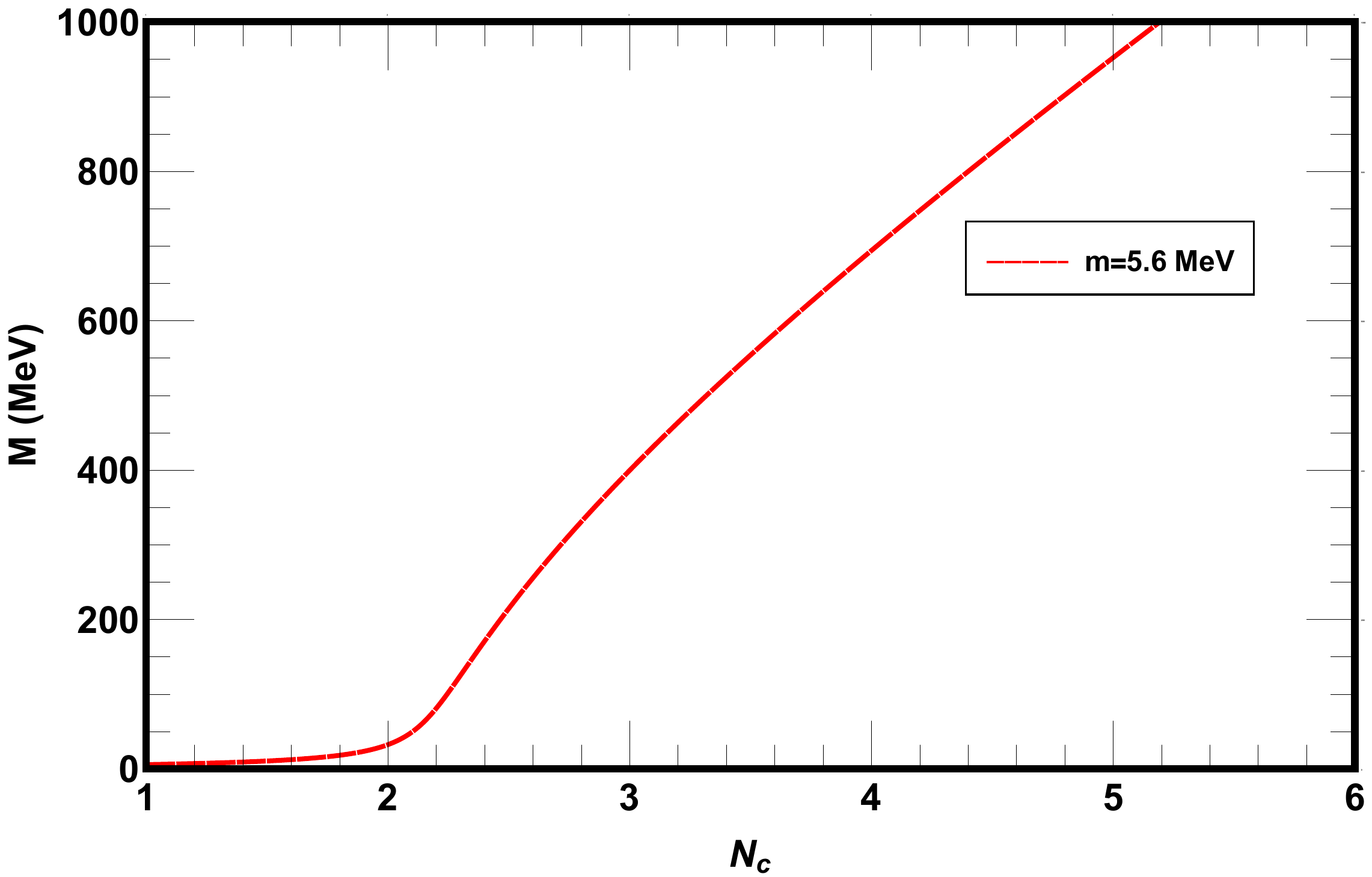}
\caption{Behavior of the dynamical mass as a function of
various number of colors $N_c$ for two flavors $N_f=2$.\label{Fig1}}
\end{center}
\end{figure}

\begin{figure}[h!]
\begin{center}
\includegraphics[scale=0.4]{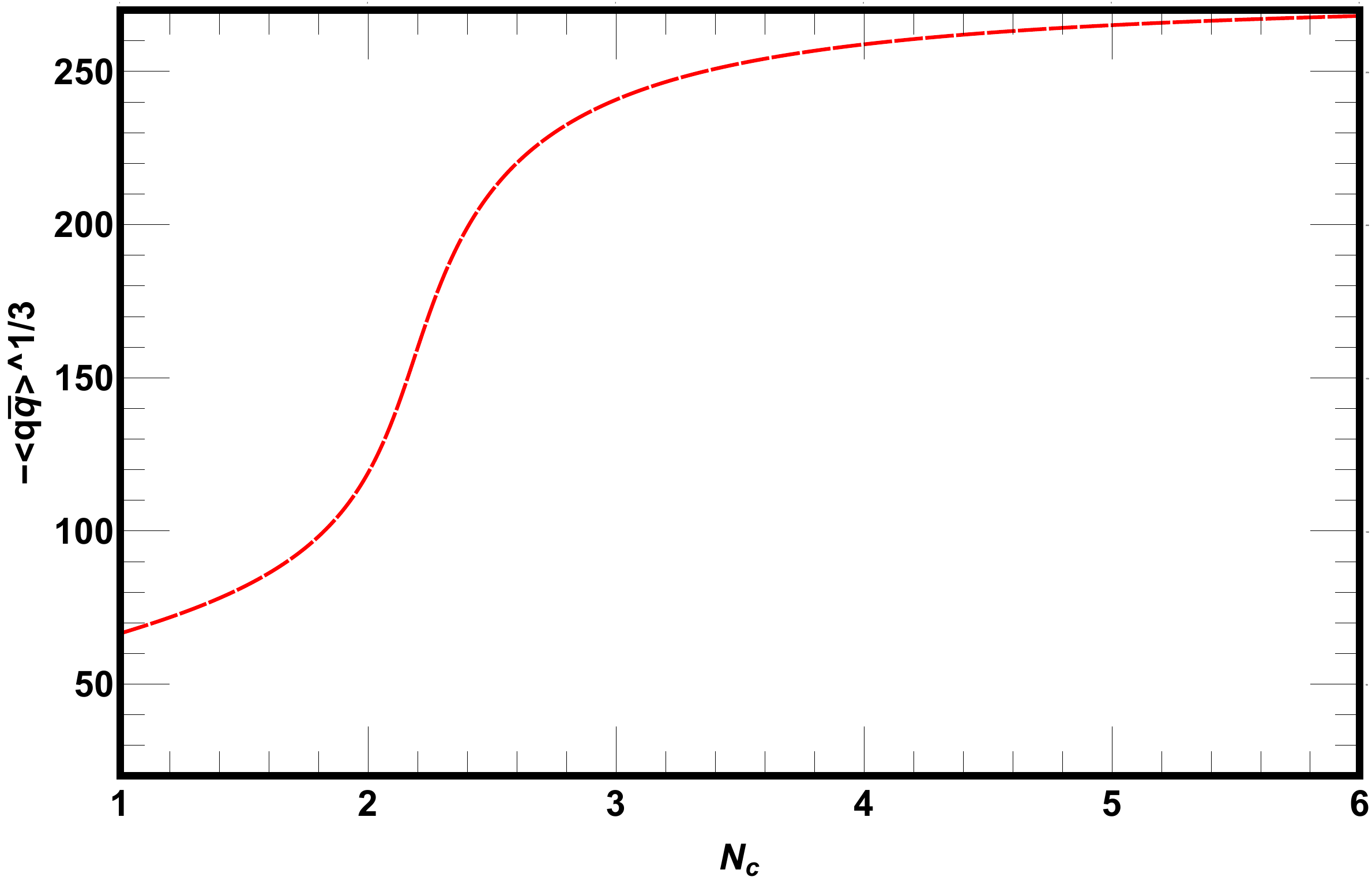}
\caption{Quark-antiquark condensate plotted as a function of number of colors $N_c$ for $N_f=2$.\label{Fig2}}
\end{center}
\end{figure}

\begin{figure}[h!]
\begin{center}
\includegraphics[scale=0.4]{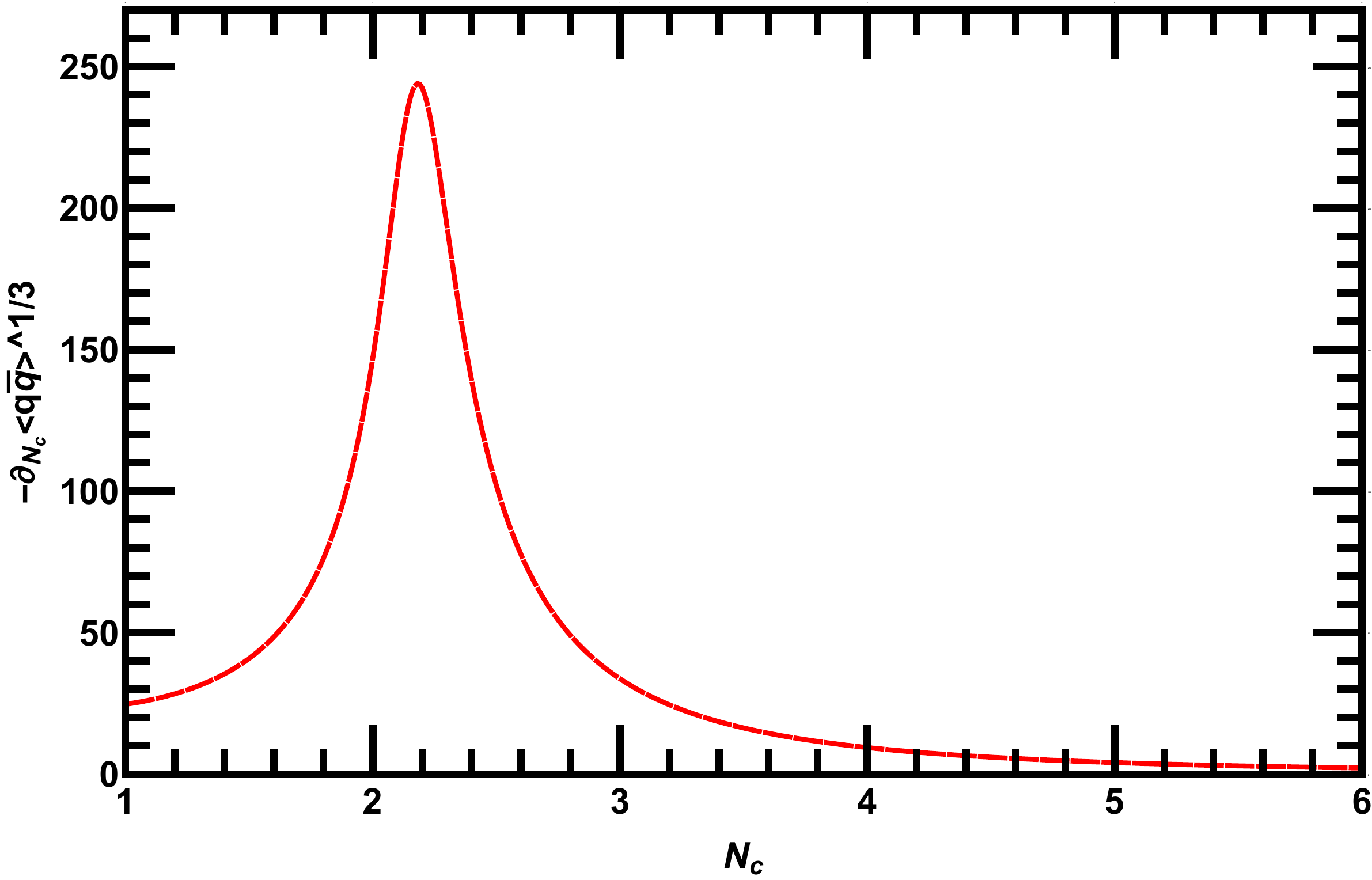}
\caption{ Color-gradient of  quark-antiquark condensate for various $N_c$, for $N_f=2$. The critical number of colors $N^{c}_c\approx 2.2$
for chiral symmetry breaking.
 \label{Fig3}}
\end{center}
\end{figure}

In this section, we numerically solve the gap equation~Eq.~(\ref{NJL16}),
with a particular choice of the parameters 
i.e., $ \Lambda = 587.9 $ MeV, $G = 2.44/ \Lambda^2$ and $m = 5.6$ MeV,
which were fitted to reproduced the pions mass in the NJL model, used in~\cite{buballa2005njl}. The solution of our  gap equation with modified 
color-flavor dependence effective coupling~Eq.~(\ref{NJL9}), for two light flavors $N_f = 2$ (i.e., up and down),
and for $Nc = 3$, yields  the dynamical mass $M = 399$
MeV. We calculated the value corresponding  quark-antiquark condensate from Eq.~(\ref{NJL17}), which is in this case $-\langle \bar{q} q\rangle^{1/3}=250$ MeV.  In this scenario, our results are consistent with that  obtained in ~\cite{buballa2005njl}, for    
$N_f = 2$ and $N_c = 3$. 
Next, we needs to solve the gap equation Eq.~(\ref{NJL16}), for
various number of colors $N_c$ and  flavors $N_f$. Initially, we solve the gap
equation for fixed flavors  $N_f = 2$, but for the various number of colors $N_c$ and plotted
the dynamically generated mass in Fig.~\ref{Fig1}, as a function of $N_c$.
We find that the dynamical chiral symmetry is partially broken when the number of colors $N_c$ exceeds a critical value $N_c=N^{c}_{c}$  and remains broken for larger values of $N_c$. The corresponding quark-antiquark condensate $-\langle \bar{q} q\rangle^{1/3}$ is shown in the Fig.~\ref{Fig2}.
The critical number of colors $N^{c}_c$ is obtained from the inflection point of the color-gradient of the quark-antiquark condensate $-\partial_{N_c}\langle \bar{q} q\rangle^{1/3}$ and is depicted in the Fig.~\ref{Fig3}.
It clear from the Fig.~\ref{Fig3} that
the dynamical chiral symmetry is partially broken above $N^{c}_{c}\approx2.2$, and thus, consistent with the  predicted value, obtained in Ref.~\cite{Ahmad:2020jzn}.

Next, we consider $N_c=3$ and solve the gap equation Eq.~\ref{NJL16}, for various number of flavors $N_f$, as shown in the Fig.~\ref{Fig3}. The plot demonstrate
that the dynamically generated mass  monotonically decreases with the increase of $N_f$ until it reaches to a critical
value $N^{c}_{f}$, where the dynamical mass vanishes and only
bare quark mass survives and thus, the chiral symmetry is partially restored above $N^{c}_f$. In Fig.~\ref{Fig5}, we plotted the  quark-antiquark condensate $-\langle \bar{q} q\rangle^{1/3}$ as a function of $N_f$ and its flavor-gradient $-\partial_{N_f}\langle \bar{q} q\rangle^{1/3}$ is depicted in the  Fig.~\ref{Fig6}. Thus,
we obtained the critical number of 
flavors $N^{c}_{f}\approx8$ from the inflection point of the flavor-gradient of the quark-anitquark condensate $-\partial_{N_f}\langle \bar{q} q\rangle^{1/3}$. Our results in this scenario
agree with that obtained in Ref.~\cite{Ahmad:2020jzn}. In   Fig.~\ref{Fig7}, we show the  variation of  critical number of flavors $N^{c}_{f}$ versus critical number of colors $N^{c}_{c}$. The  Fig.~\ref{Fig7}, clearly demonstrate that the  critical value $N^{c}_{f}$ enhances with as 
the $N^{c}_{c}$ shifted toward its larger values, and thus both the parameters opposes the effect of  each other. For lower $N^{c}_{c}$, the  $N^{c}_{f}$ has a smaller values while for higher values of  $N^{c}_{c}$, the  $N^{c}_{f}$
enhances. As, we discussed before that  the number
colors $N_c$ anti-screens the strong interactions  while  $N^{c}_{f}$ screens  them and is  readily confirmed from the Fig.~\ref{Fig7}.  For clear understanding, we have tabulated  some data for the variation of 
critical number of  flavors $N^{c}_{f}$ with various  critical number of colors $N^{c}_{c}$ in
the Tab.~\ref{NCNF}.
\begin{table}[h!]
\begin{center}
\caption{The variation of critical Number of 
flavors $N^{c}_{f}$ with the critical number of colors $N^{c}_{c}$.\label{NCNF}}
\begin{tabular}{|c|c|c|c|c|c|c|c|c|c|}
\hline
\textbf{$N^{c}_{c}$} &2.2  &3 &4 &5 &6 &7 &8 & 9&10 \\[6pt]
\hline
\textbf{Critical $ N^{c}_{f}$}& 3  &8 &9.5 &10.3 &10.7 &11 &11.2 & 11.3&11.4 \\[6pt]
\hline
\end{tabular}
\end{center}
\end{table}
\begin{figure}[h!]
\begin{center}
\includegraphics[scale=0.4]{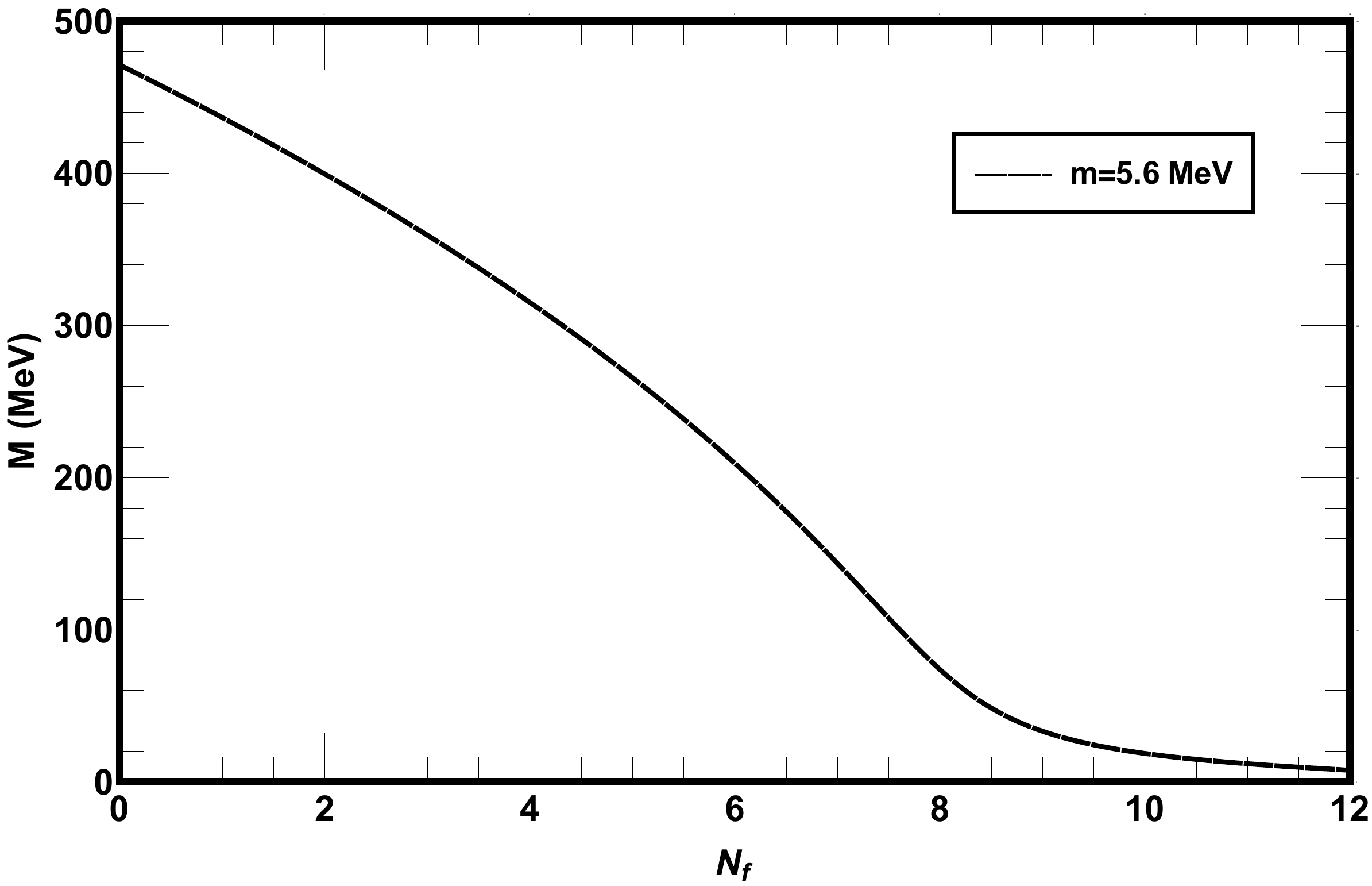}
\caption{Behavior of the dynamical mass  as a function of various number of flavor $N_f$ with fixed number of colors $N_c=3$. The dynamical
mass monotonically decreases with  increasing $N_f$.\label{Fig4}}
\end{center}
\end{figure}
\begin{figure}[h!]
\begin{center}
\includegraphics[scale=0.4]{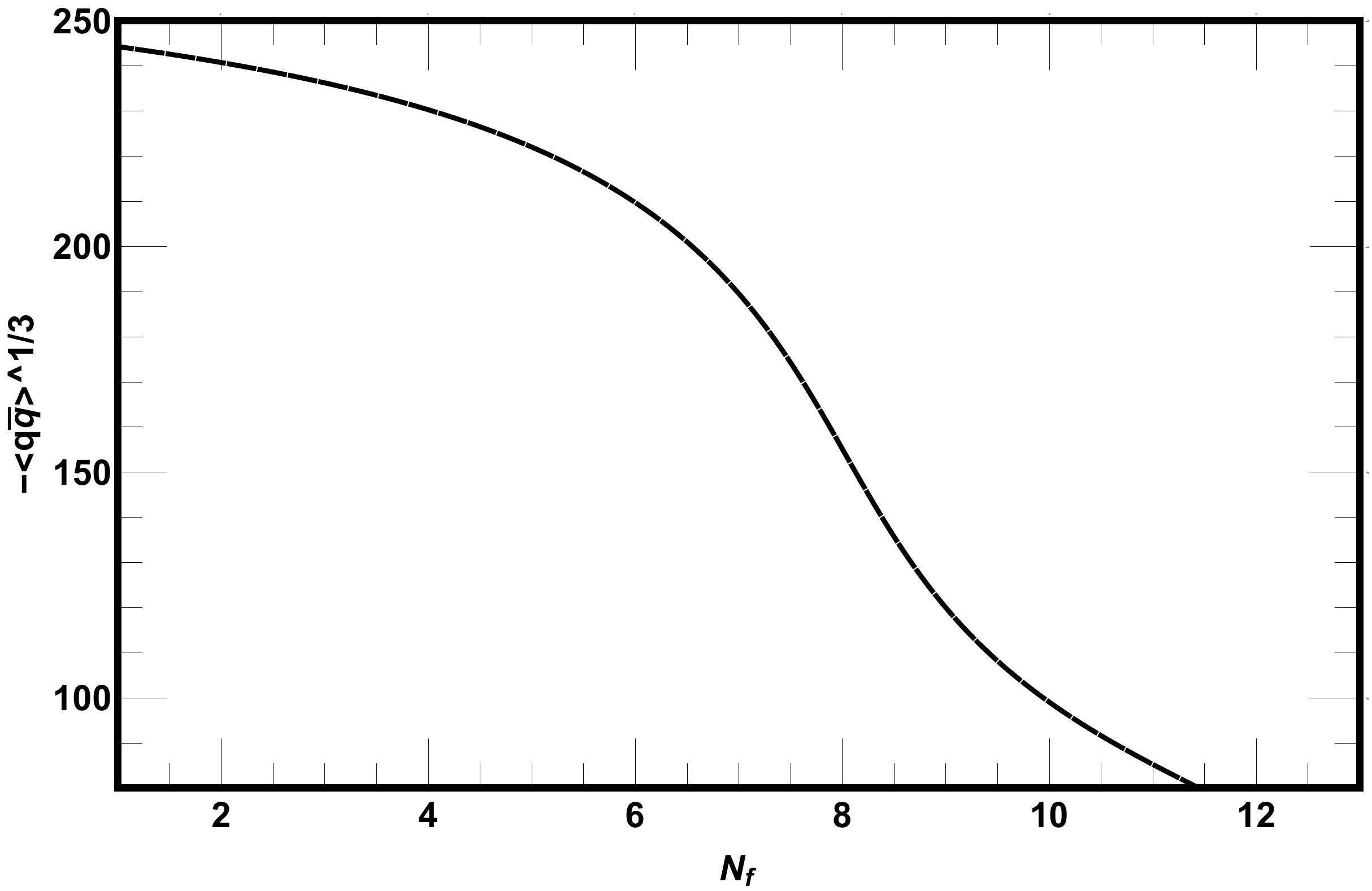}
\caption{Quark-antiquark condensate as a function of number of flavors $N_f$ with fixed colors $N_c=3$. \label{Fig5}}
\end{center}
\end{figure}

\begin{figure}[h!]
\begin{center}
\includegraphics[scale=0.4]{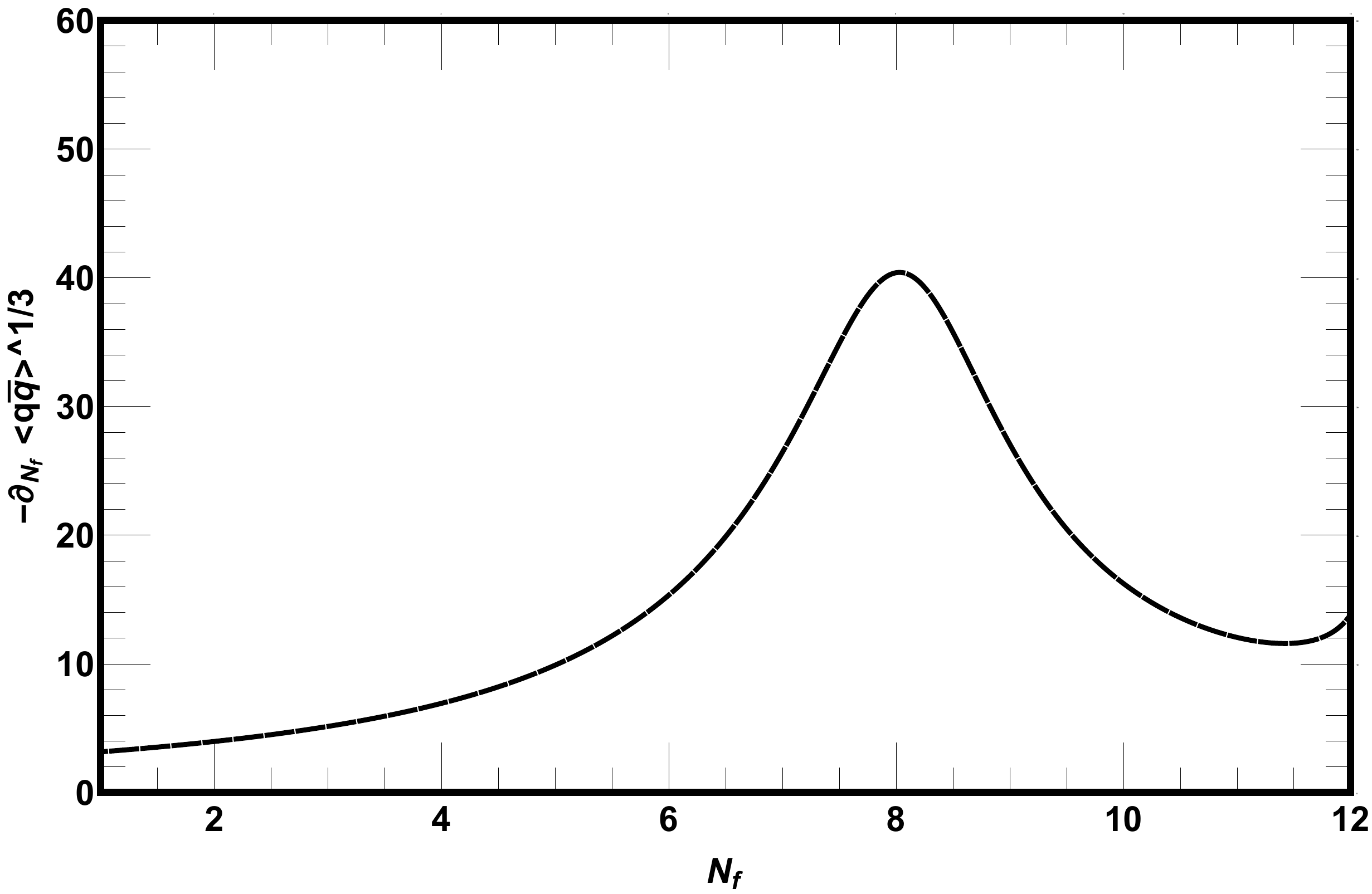}
\caption{Flavor-gradient of the quark-antiquark condensate as
a function of flavors $N_f$. The inflection point of the gradient
represent the critical number of flavors  for the dynamical chiral symmetry restoration and is  at $N^{c}_{f}\approx8$.\label{Fig6}}
\end{center}
\end{figure}

\begin{figure}[h!]
\begin{center}
\includegraphics[scale=0.4]{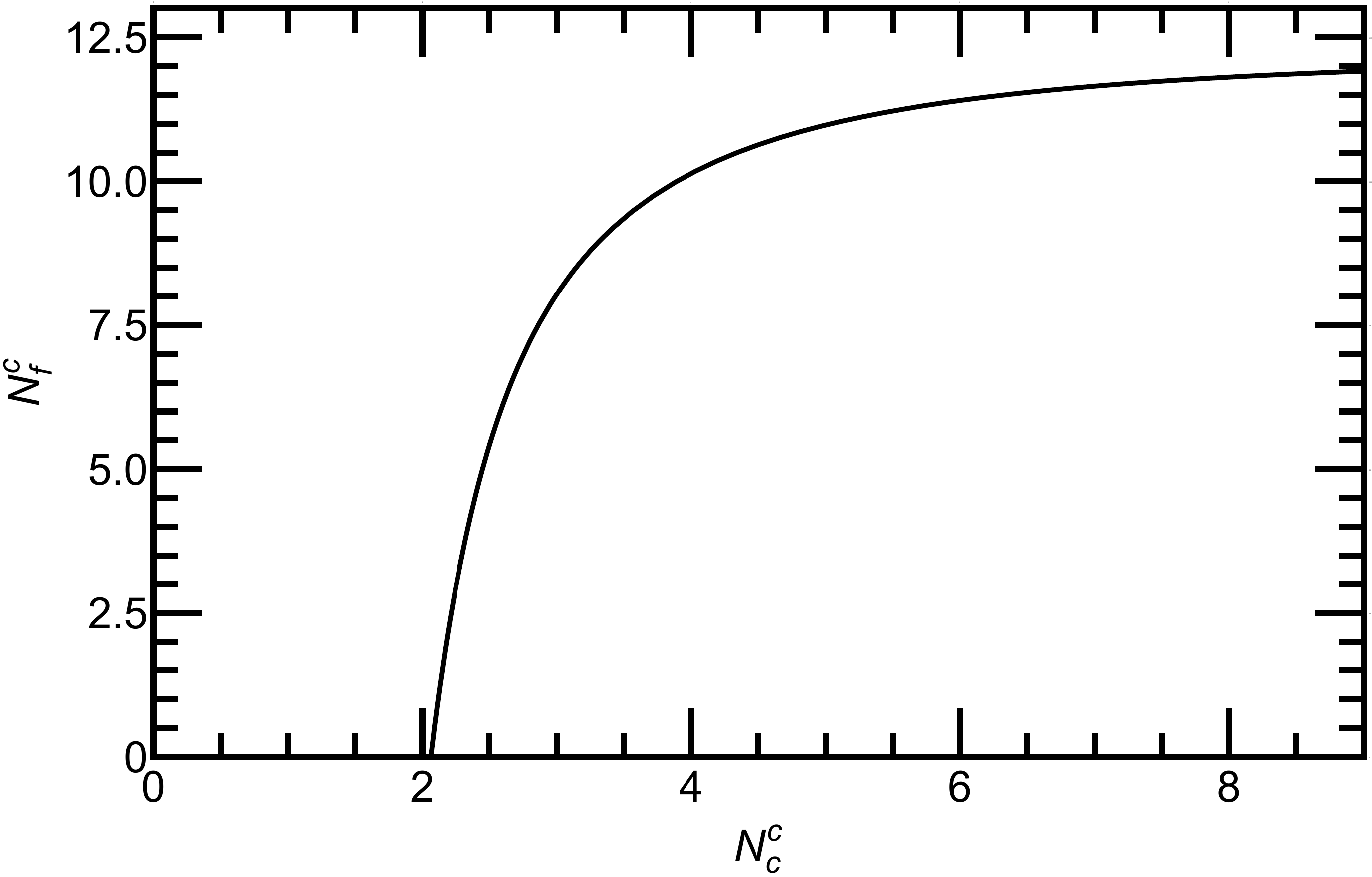}
\caption{Behavior of the critical number of flavors $N^{c}_{f}$ versus critical
number of colors $N^{c}_{c}$  
for chiral symmetry breaking/restoration.  \label{Fig7}}
\end{center}
\end{figure}
In the next section, we shall discuss the dynamical chiral symmetry breaking and its restoration with colors  $N_c$  and flavors $N_f$ and at finite temperature $T$.
\section{Dynamical chiral symmetry breaking with $N_c$, $N_f$ and at finite $T$}\label{section-IV}
\begin{figure}[h!]
\begin{center}
\includegraphics[scale=0.4]{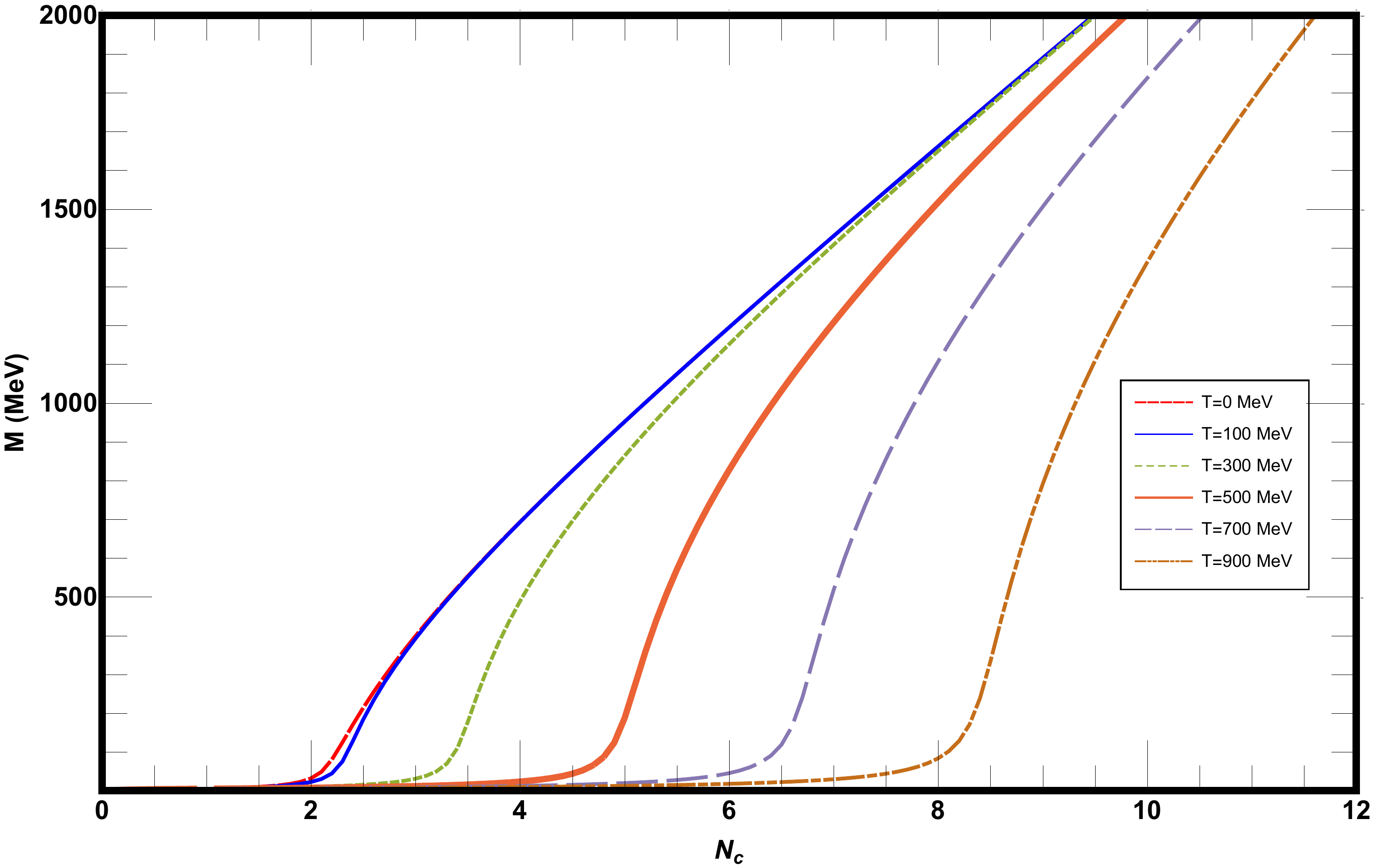}
\caption{Behavior of  dynamical mass as a function of $N_c$ for two flavors$N_f = 2$, and for 
for various  temperatures $T$. The higher the temperature $T$ the large number of critical number of colors  $N^{c}_{c}$ required for the dynamical symmetry breaking.\label{Fig8}}
\end{center}
\end{figure}

\begin{figure}[h!]
\begin{center}
\includegraphics[scale=0.4]{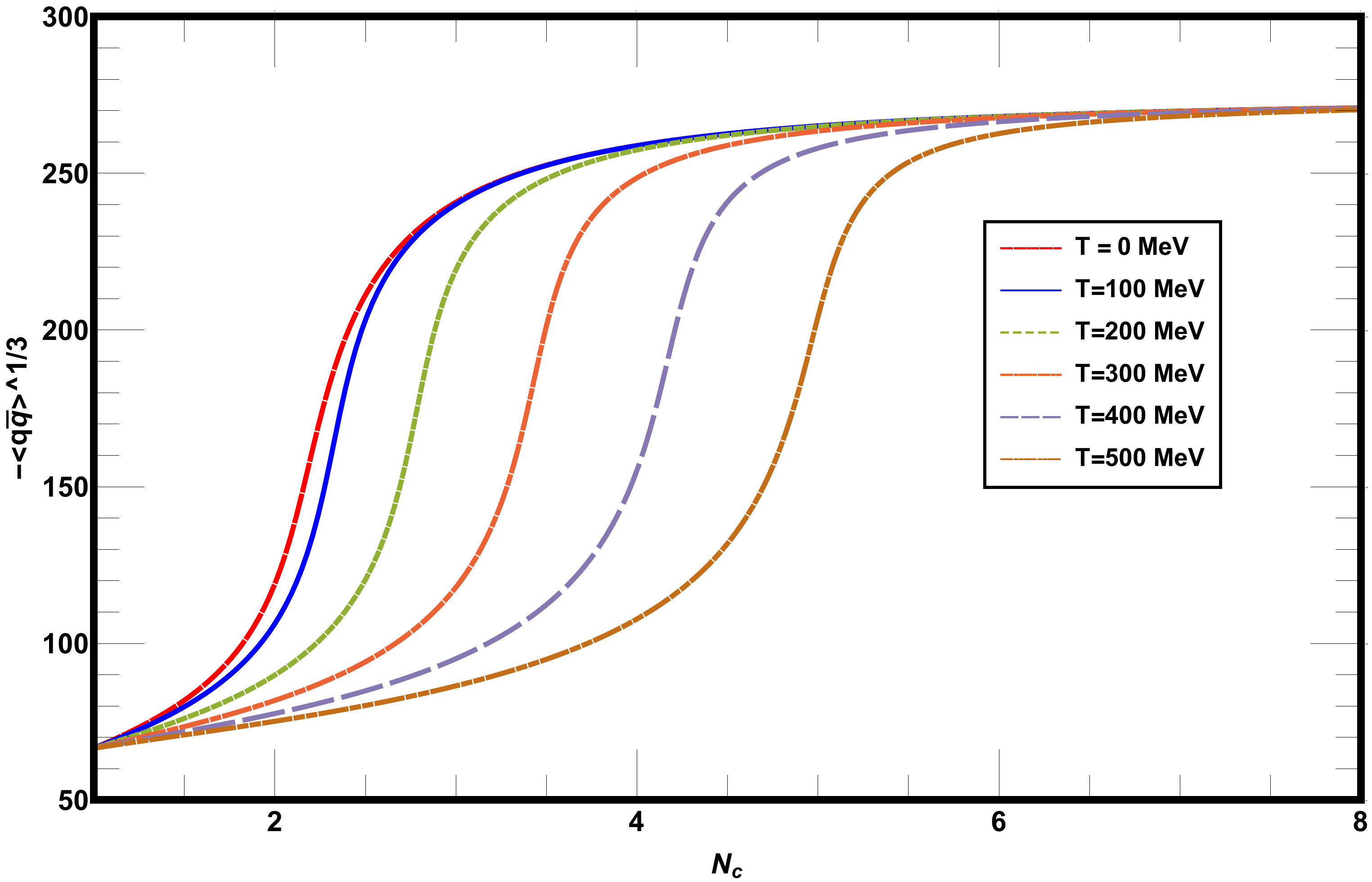}
\caption{Quark-aniquark condensate as a function of $N_c$, 
for various $T$ and for two flavors  $N_f = 2$.\label{Fig9}}
\end{center}
\end{figure}

\begin{figure}[h!]
\begin{center}
\includegraphics[scale=0.4]{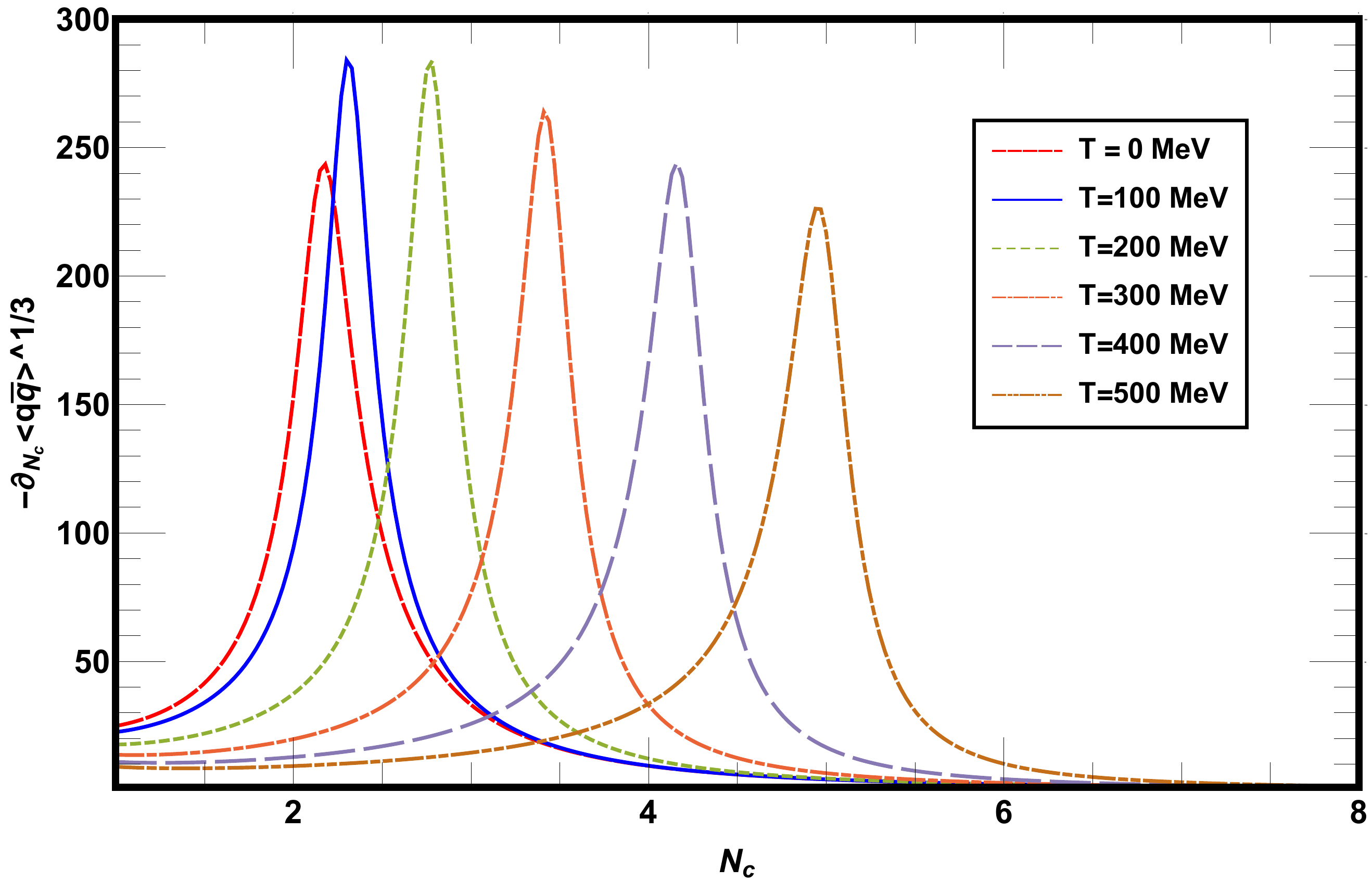}
\caption{The color-gradient of quark-antiquark condensate for various temperatures $T$. The inflection points of the gradient shifted toward their larger values of  critical  number of colors $N^{c}_{c}$ upon increasing the temperature $T$.\label{Fig10}}
\end{center}
\end{figure}

In this section,  we numerically  solve the gap equation Eq.~(\ref{NJL21})  to  understand the dynamical symmetry breaking and restoration  for  colors $Nc$, 
and  flavors $N_f$ and at finite temperature $T$.
Initially, we set $N_f=2$  and plotted the dynamical mass in Fig.~\ref{Fig8} as a function $N_c$, for various $T$ and the corresponding condensate depicted in Fig.~\ref{Fig9}. We see that, upon increasing $T$, the dynamical symmetry is broken above the critical value of colors $N^{c}_c$.
We determine the  $N^{c}_{c}$, for various values of $T$ from the inflection points of the  color-gradient $-\partial_{N_c}\langle \bar{q} q\rangle^{1/3}$,as shown in the Fig.~\ref{Fig10}. The
peaks of the gradients shifts toward larger values of $N^{c}_c$
as we increase $T$. This is because, on one
hand $N_c$ strengthen the interactions while  $T$ suppresses them and as a result, we needed large   number of critical colors $N^{c}_c$ for the dynamical chiral symmetry breaking.
\begin{figure}[h!]
\begin{center}
\includegraphics[scale=0.4]{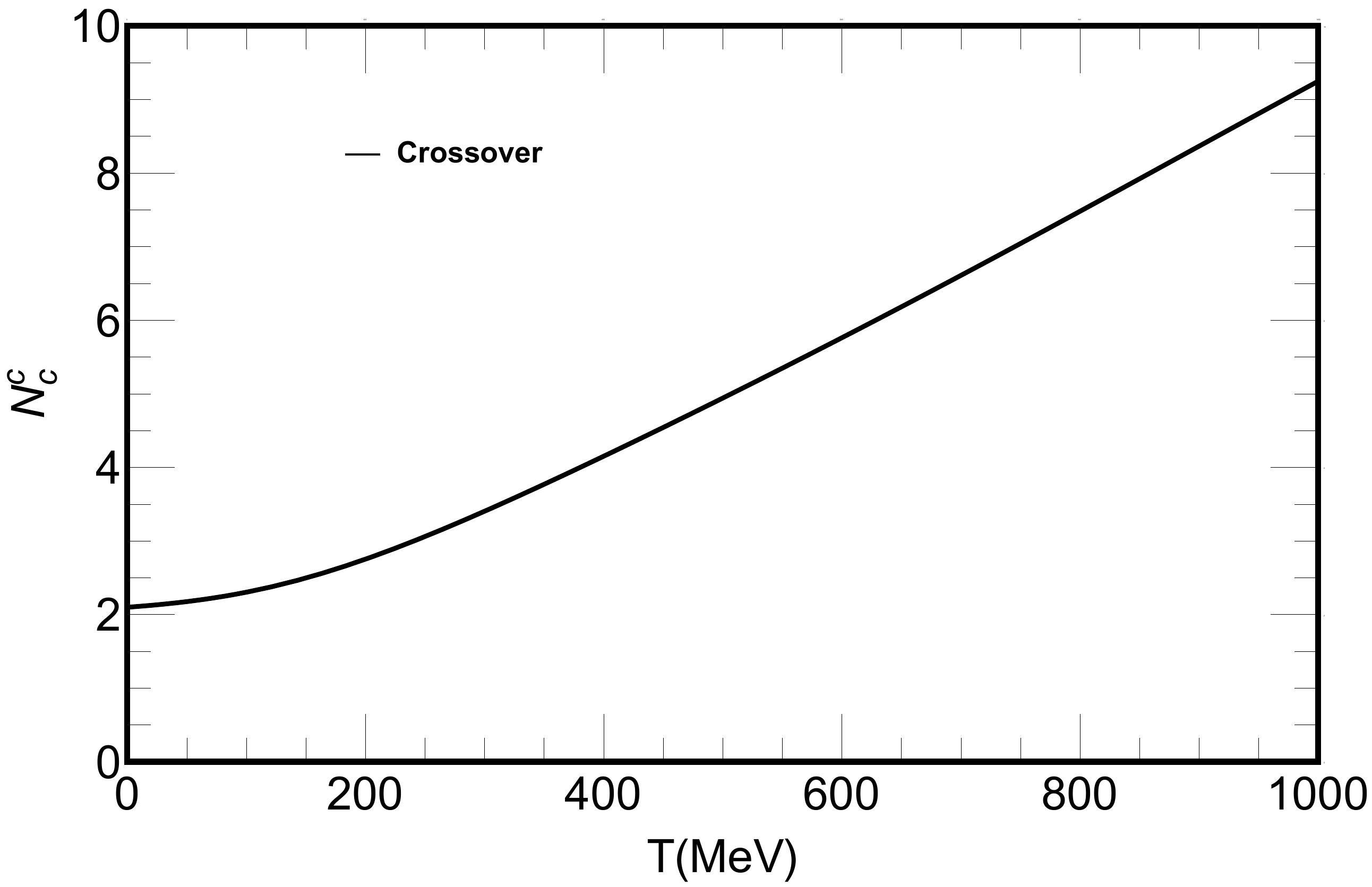}
\caption{The phase diagram for the dynamical chiral symmetry breaking/restoration for critical number of colors $N^{c}_{c}$ versus critical temperature $T=T_c$. \label{Fig11}}
\end{center}
\end{figure}
Next, we set $N_c = 3$, and plotted the dynamical mass as a function of $N_f$
, for different values of $T$
in Fig.~\ref{Fig12} and  the  corresponding quark-antiquark condensate  in  the Fig.~\ref{Fig13}. In this case, we found that the dynamical
mass as a function $N_f$ suppresses as  the $T$ increases. This
is because both  the parameters, $N_f$ and $T$ screens the strong interactions,
and  thus, the dynamical chiral symmetry breaks even 
for smaller values of $N_f$. We thus, obtained the critical  number of flavors
$N_f$, for different $T$ from the inflection points of the 
flavor-gradient $-\partial_{N_f}\langle \bar{q} q\rangle^{1/3}$, as depicted in the Fig.~\ref{Fig14}. 
We noted that the peaks are shifted toward their lower $N^{c}_f$ values. We thus, observe that  $N^{c}_{f}$ monotonically decreases as $T$ increases, as demonstrated in the Fig.~ \ref{Fig15}. This means that in the presence of heat both, less number of critical  flavors $N^{c}_{f}$
required for the chiral symmetry breaking and  restoration. The nature of the transition at each temperature $T=T_c$ is observed to be cross-over. 
\begin{figure}[h!]
\begin{center}
\includegraphics[scale=0.4]{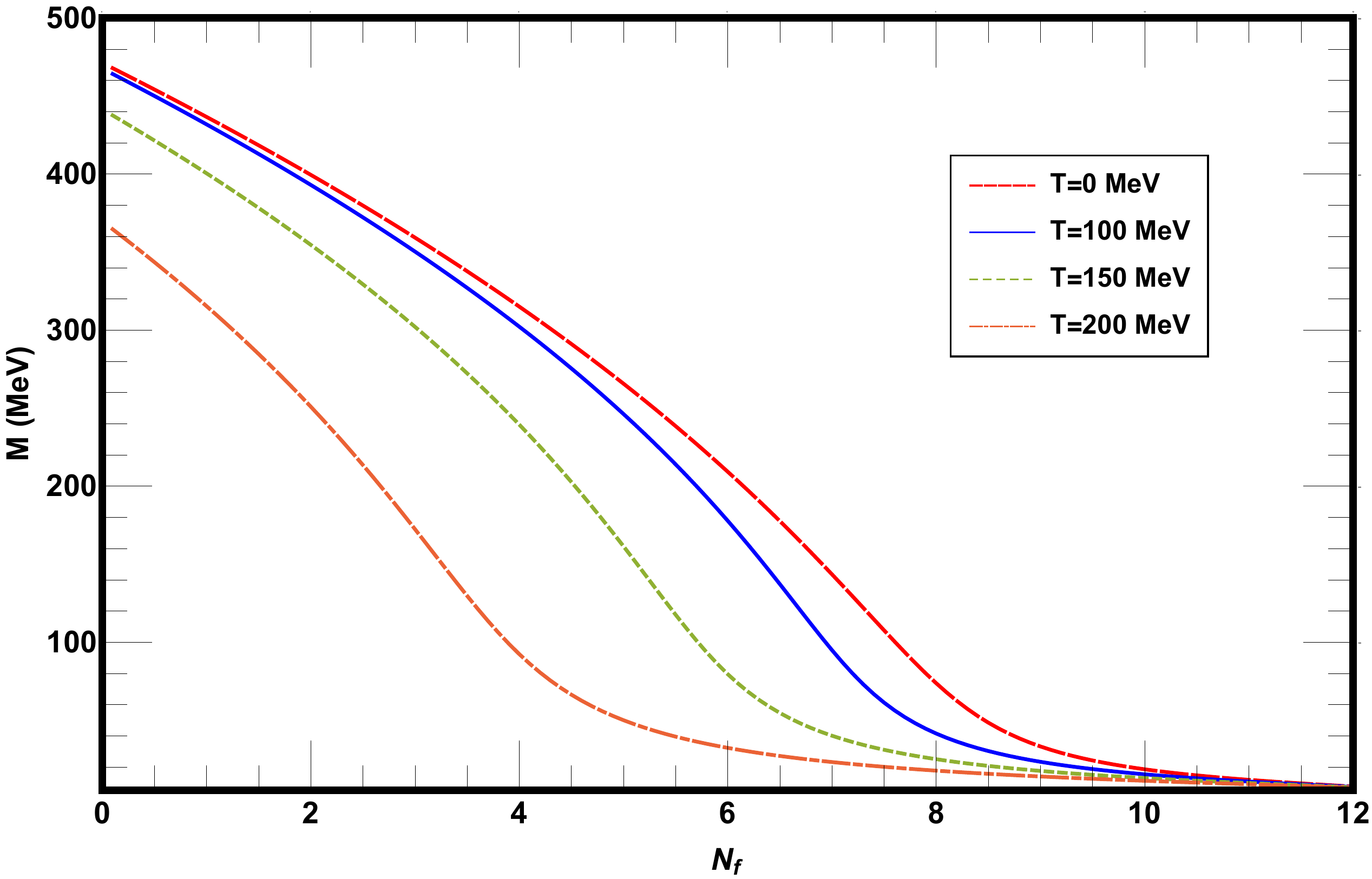}
\caption{The behavior of dynamical mass as a function of number of flavors $N_f$, for  fixed $N_c=3$ and for various  temperature $T$. Upon increasing the temperature $T$ the dynamical mass suppresses.\label{Fig12}}
\end{center}
\end{figure}
\begin{figure}[h!]
\begin{center}
\includegraphics[scale=0.4]{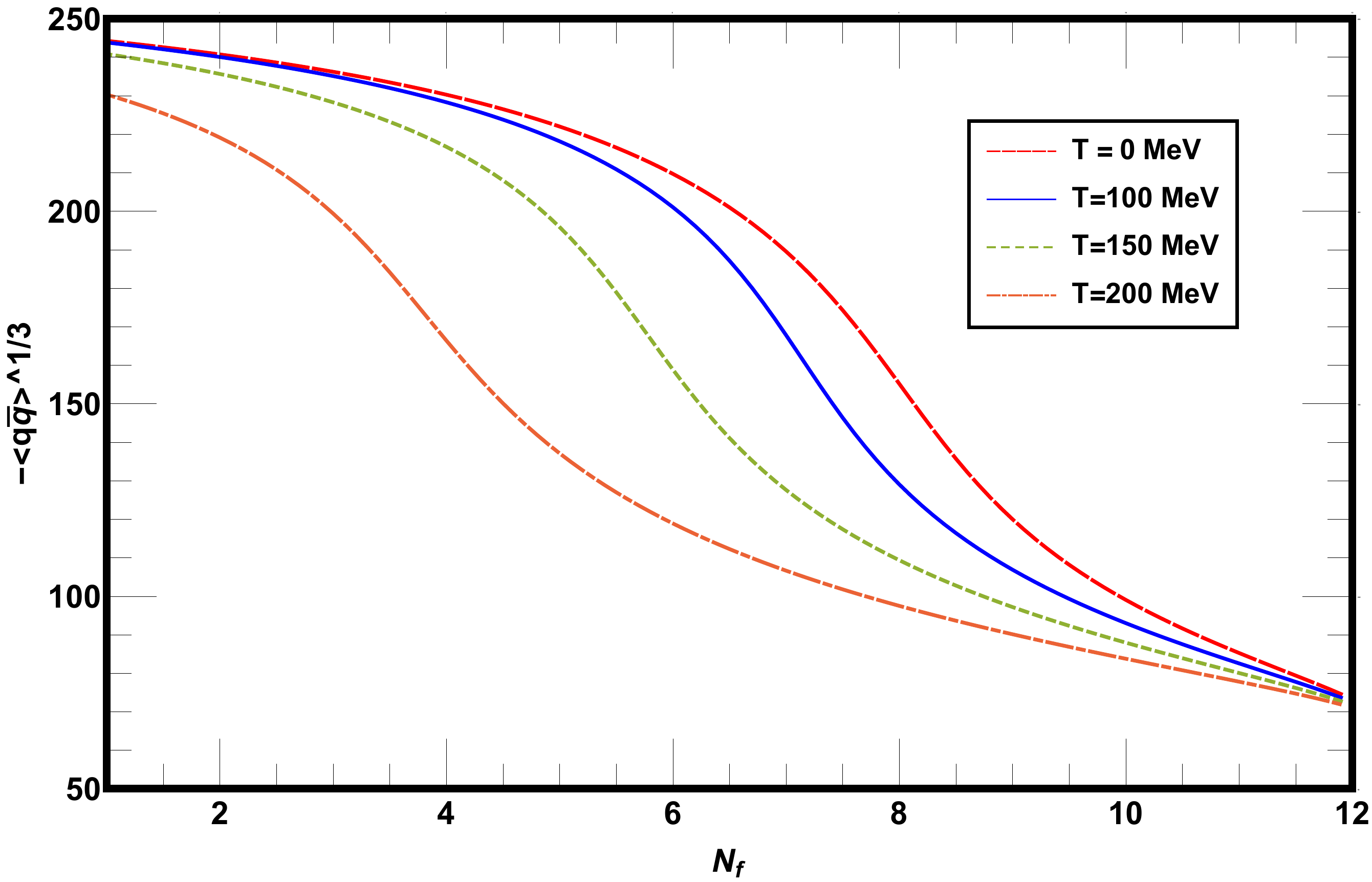}
\caption{The behavior of quark-antiquark condensate as a function of number of flavors $N_f$, for  fixed $N_c=3$ and for various  temperature $T$. \label{Fig13}}
\end{center}
\end{figure}
\begin{figure}[h!]
\begin{center}
\includegraphics[scale=0.4]{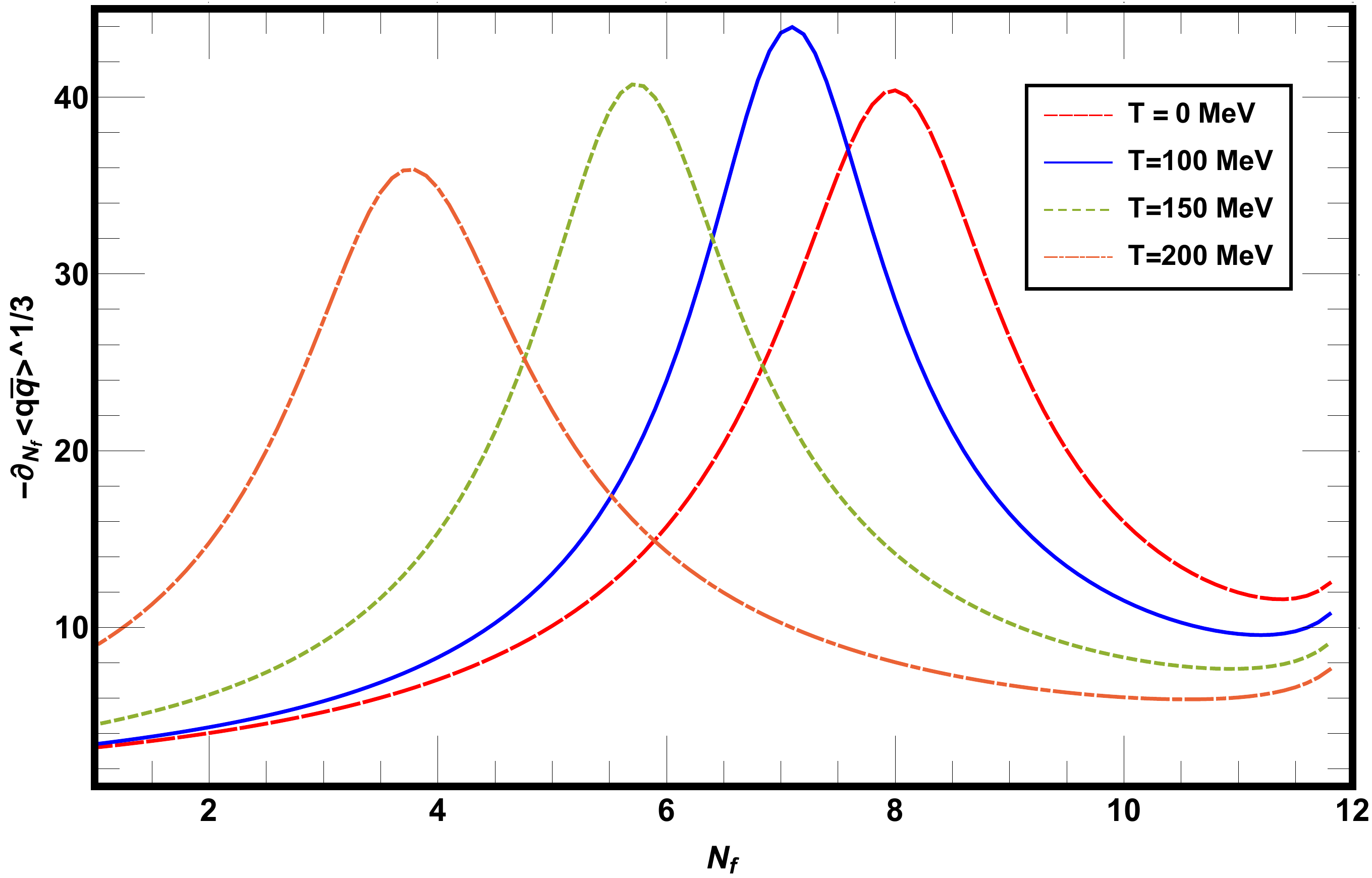}
\caption{The flavor-gradient of the condensate at different temperature $T$, which shows that upon increasing the temperature the inflection points shifts toward their lower $N^{c}_{f}$ values.\label{Fig14}}
\end{center}
\end{figure}
\begin{figure}[h!]
\begin{center}
\includegraphics[scale=0.4]{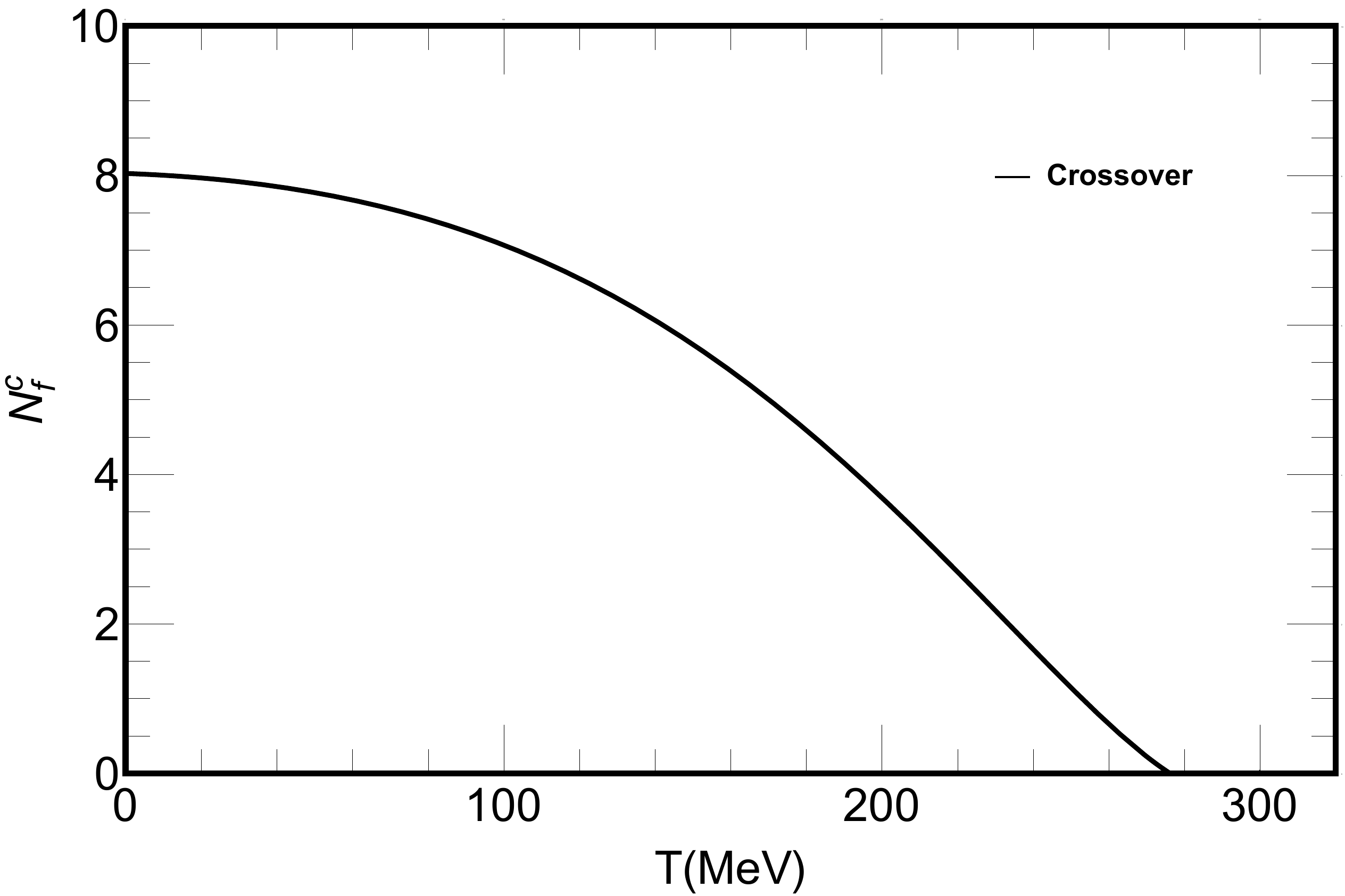}
\caption{The phase diagram for the dynamical chiral symmetry breaking/restoration for critical number of flavors $N^{c}_{f}$ versus critical temperature $T=T_c$.\label{Fig15}}
\end{center}
\end{figure}
In the next section, we shall investigate the behavior of chiral symmetry breaking and restoration at a finite chemical potential $\mu$ upon increasing both the number of colors $N_c$ and flavors $N_f$.

\section{Dynamical Chiral Symmetry Breaking for  $N_f$, $N_c$ and at finite $\mu$}\label{section-V}
\begin{figure}[h!]
\begin{center}
\includegraphics[scale=0.4]{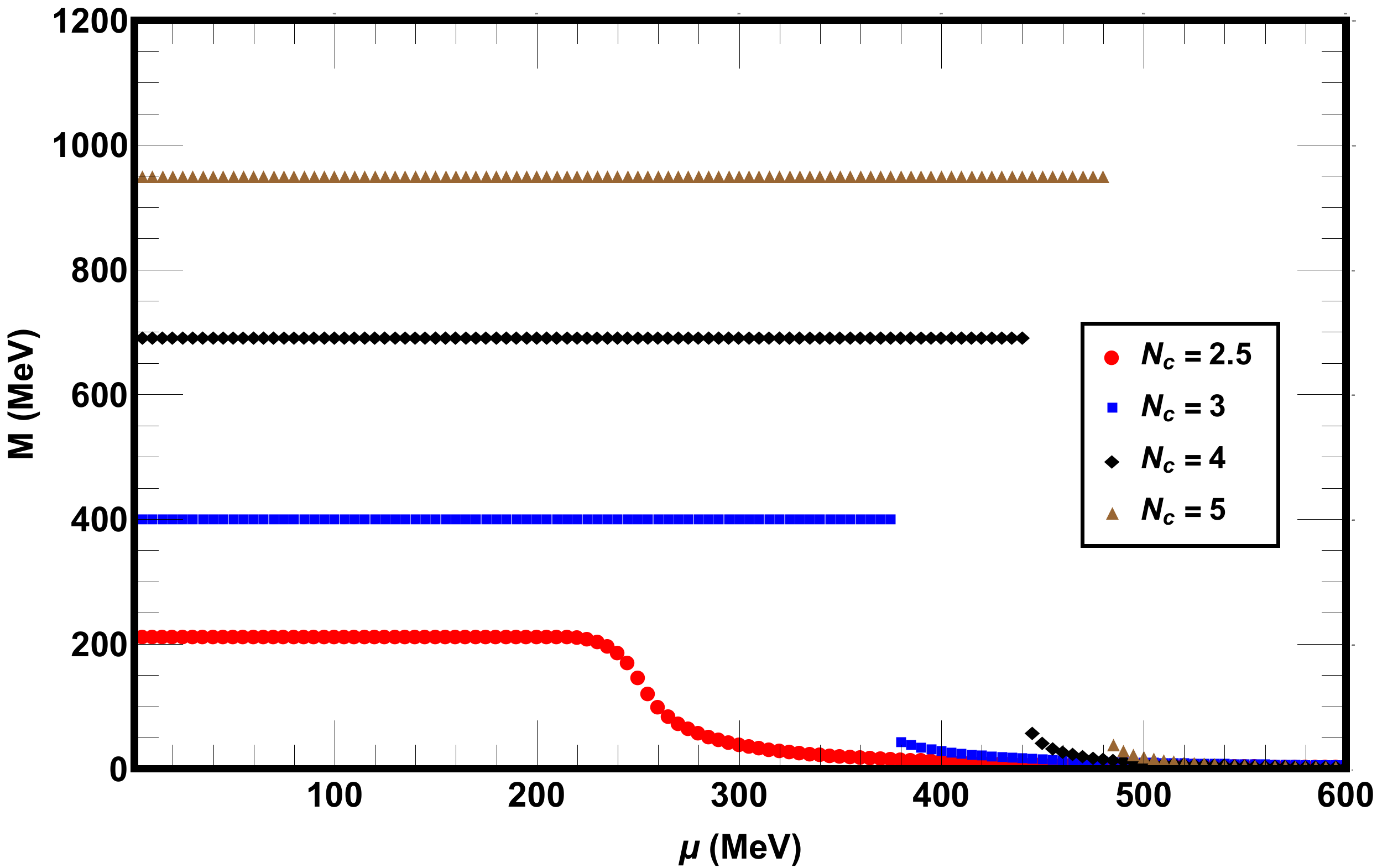}
\caption{ Behavior of the dynamical  quark mass as a function of quark chemical potential $\mu$ for various number of colors $N_c$. The Plot show that the dynamical chiral symmetry restored above some critical $\mu_c$, for each $N_c$.\label{Fig16}}
\end{center}
\end{figure}

\begin{figure}[h!]
\begin{center}
\includegraphics[scale=0.4]{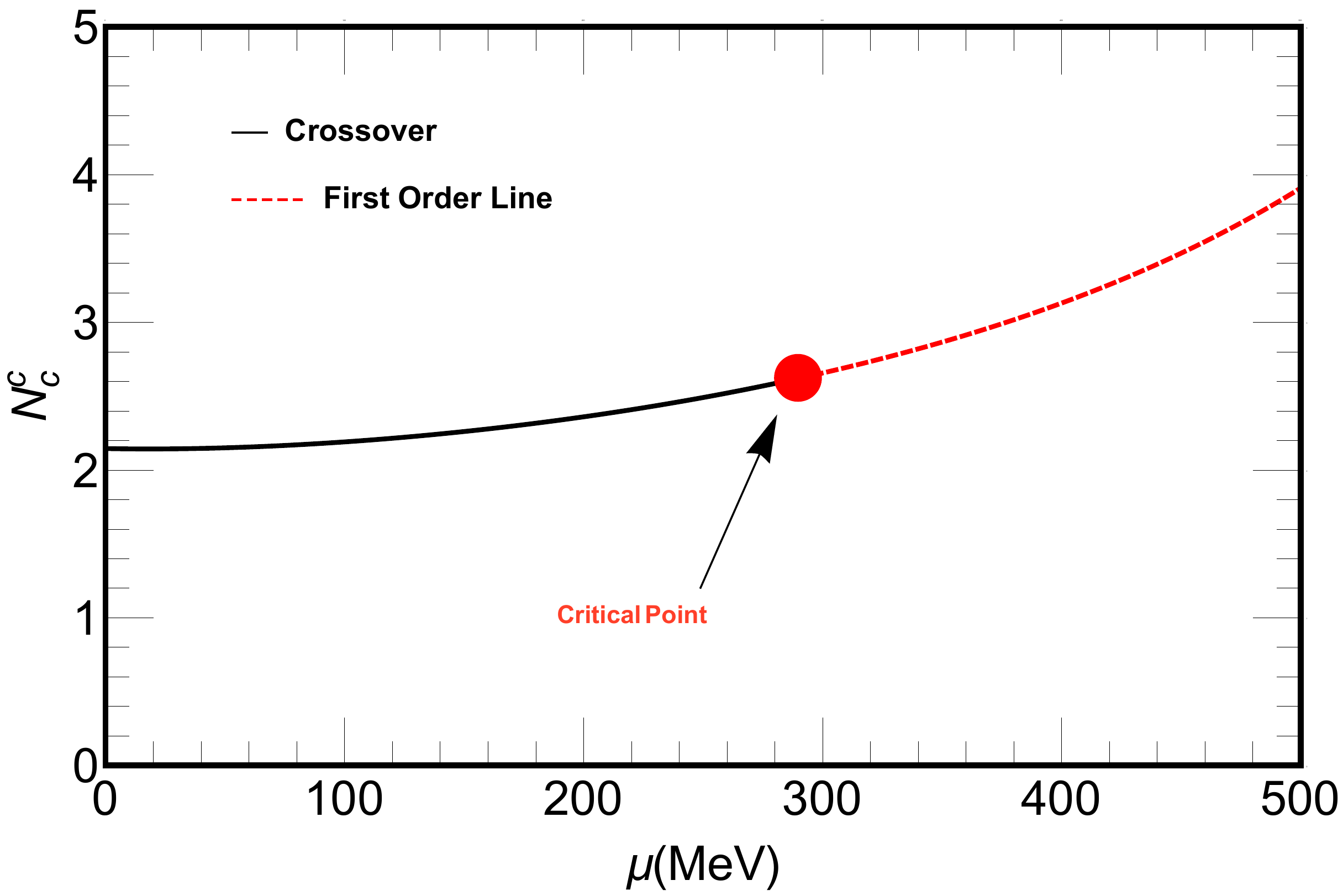}
\caption{The phase diagram for the dynamical chiral symmetry breaking/restoration for critical number of colors $N^{c}_{c}$ versus critical chemical potential $\mu=\mu_c$. The nature of the phase transition is smooth cross-over until the  critical  endpoint $(N_{c}^{c}\approx2.5, \mu^{c}_c\approx 290$ MeV), and above this point the transition changes to  the first order.\label{Fig17}}
\end{center}
\end{figure}

\begin{figure}[h!]
\begin{center}
\includegraphics[scale=0.4]{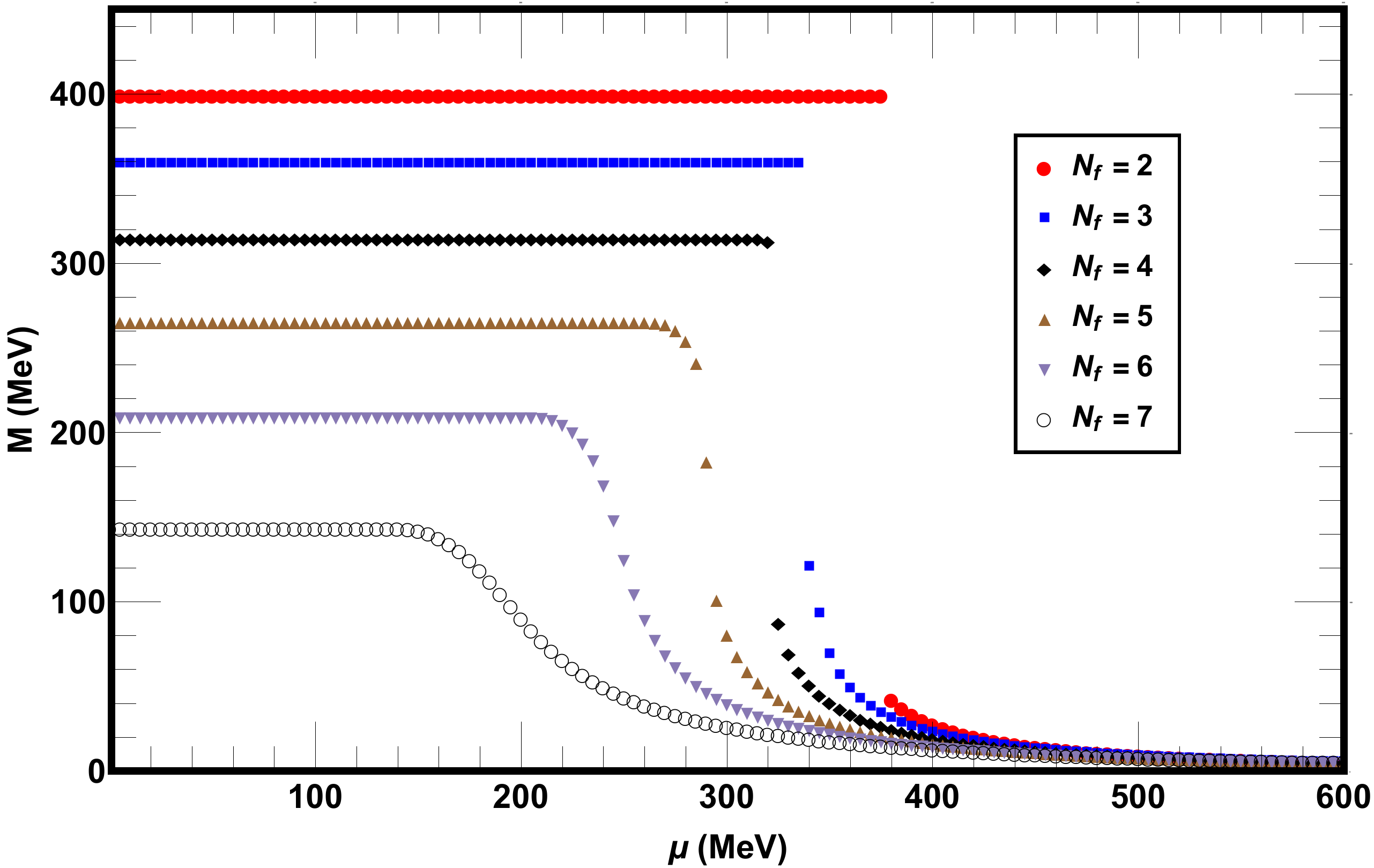}
\caption{Behavior of the dynamical  quark mass as a function of quark chemical potential $\mu$ for various number of flavors $N_f$. The plot show that the dynamical chiral symmetry restored above some critical $\mu_c$, for each $N_c$.\label{Fig18}}
\end{center}
\end{figure}

\begin{figure}[h!]
\begin{center}
\includegraphics[scale=0.4]{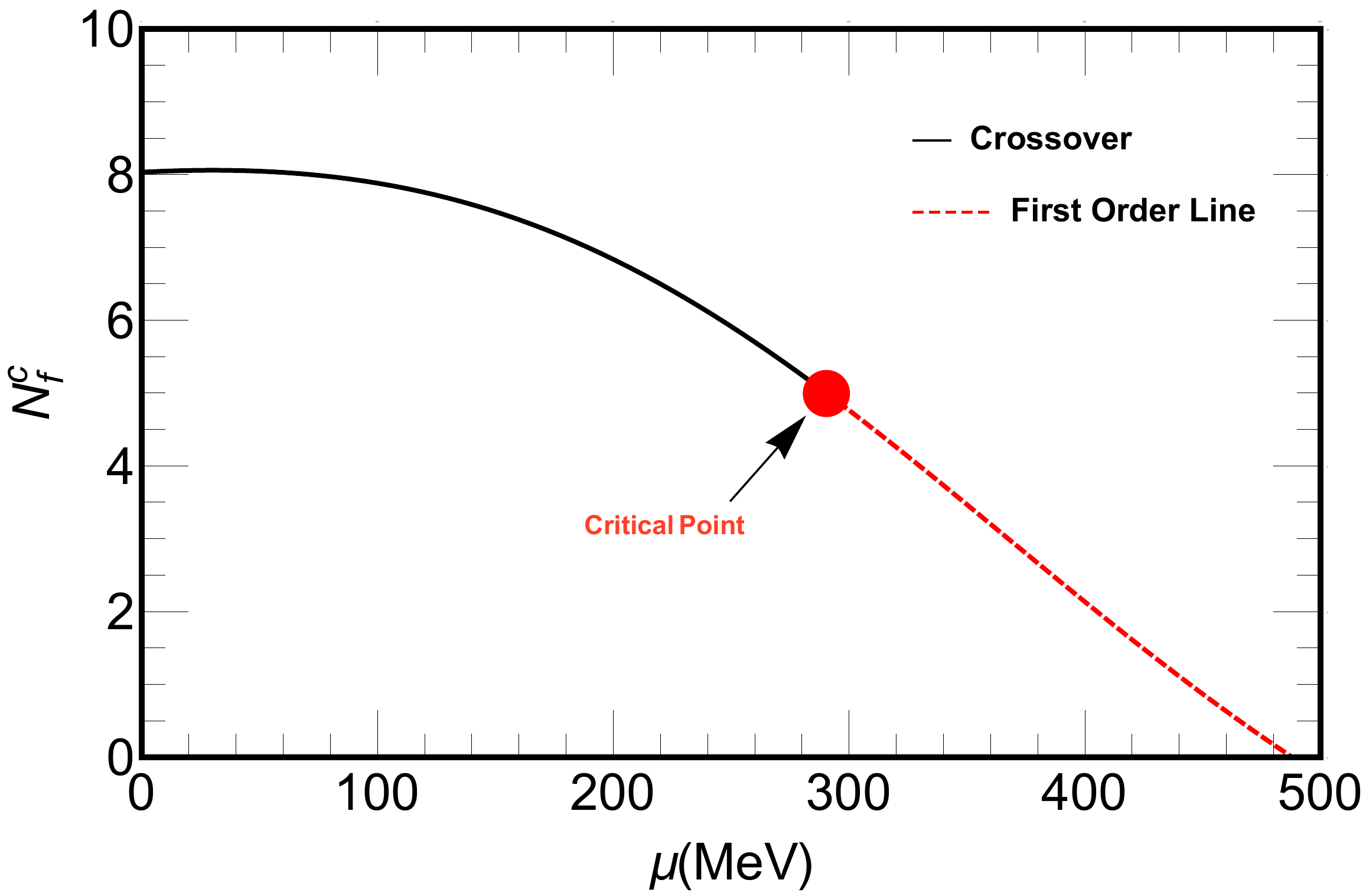}
\caption{The phase diagram for the dynamical chiral symmetry breaking/restoration for critical number of flavors $N^{c}_{f}$ versus critical chemical potential $\mu=\mu_c$. The nature of the phase transition is smooth cross-over for $N_f\geq5$,  while for $N_f\geq5$  the transition  changes to first order. Both transition are differentiated by a critical  endpoint $(N_{c}^{f}\approx5 ,~\mu^{f}_c \approx 290$ MeV).\label{Fig19}}
\end{center}
\end{figure}

In the present section, we discuss the dynamical chiral symmetry breaking and its restoration at finite quark chemical potential $\mu$,  for various $N_c$  and  $N_f$. 
The dynamical mass as a function of $\mu$ for various $N_c$ is shown in Fig.~\ref{Fig16}. 
We see that the dynamical chiral symmetry is partially restored when the chemical potential $\mu$ exceeds a  critical value $\mu_c$. The discontinuity in the dynamical mass around $\mu^{c}_c$ shows that the nature of the phase transition is of first-order, while the smooth decrease in the dynamical mass, represents the cross-over phase transition.  It means that quark chemical potential $\mu$, produces the screening effect, in contrast to the anti-screening produced by higher $N_c$.  We determined the  critical number of colors  $N^{c}_{c}$ from the inflection point of the color-gradient $-\partial_{N_c}\langle \bar{q} q\rangle^{1/3}_{\mu}$ at different $\mu$, and plotted the variation of the $N^{c}_{c}$ versus $\mu$ in the Fig.~\ref{Fig17}. 
This  plot show that with an increase in chemical potential $\mu$, the critical number of colors  $N^{c}_{c}$ for chiral symmetry breaking also increases. For example, at $\mu = 100$ MeV, the chiral symmetry breaking needs less critical  number of  colors (i.e, $N^{c}_c= 2.2$) as compared to $ \mu = 250$ MeV (i.e., $N^{c}_c= 2.5$ ). We found the cross-over phase transition  for $ N_c \leq 2.5 $,  and first order for  $ N_c > 2.5 $ differentiated  by a critical endpoint  $(N_{c}^{c}\approx2.5,~\mu^{c}_c\approx290$ MeV). 
In Fig.~\ref{Fig18}, we plotted the dynamical generated mass as a function of $\mu$, for various number of flavors $N_f$ and for fixed $N_c=3$.  The dynamical chiral symmetry is partially restored when the chemical potential reaches its critical value $\mu_c$  for various  $N_f$. However, the nature of phase transition is off first- order for $N_f\leq5$, and cross-over for $N_f\geq5$.   The dynamical mass as a function of chemical potential $\mu$  suppresses with the increase of $N_f$.  Thus, the chemical potential $\mu$ also screens the interactions in resemblance with the screening effect produced by $N_f$. 
We determined  the  critical number of flavors $N^{c}_f$  for each chemical potential $\mu$ from the inflation point of  $-\partial_{N_c}\langle \bar{q} q\rangle^{1/3}_{\mu}$, and plotted the variation of the $N^{c}_{f}$ versus critical chemical potential $\mu_c$ in the Fig.~\ref{Fig19}.  We noted that the critical number of flavors $N^{c}_{f}$ decreases with the increase of chemical potential $\mu$.  The nature of the phase transition is off first-order until the critical endpoint $(N_{c}^{f}=5, \mu^{f}_c=290$ MeV), where the transition changes to first order. 
The overall results show that at higher values of chemical potential $\mu$, we need the higher critical number of colors $N^{c}_{c}$ and the less critical number of flavors $N^{c}_{f}$  required for  the dynamical chiral symmetry breaking/restoration. 
In the next section, we investigate the simultaneous effects of temperature  $T$ and chemical potential $\mu$ on chiral phase transition with the higher number of light quark flavors and colors.

\section{ QCD Phase Diagrams for various  $N_c$ and  $N_f$ in  $T-\mu$ plane}\label{section-VI}
\begin{figure}[h!]
\begin{center}
\includegraphics[scale=0.4]{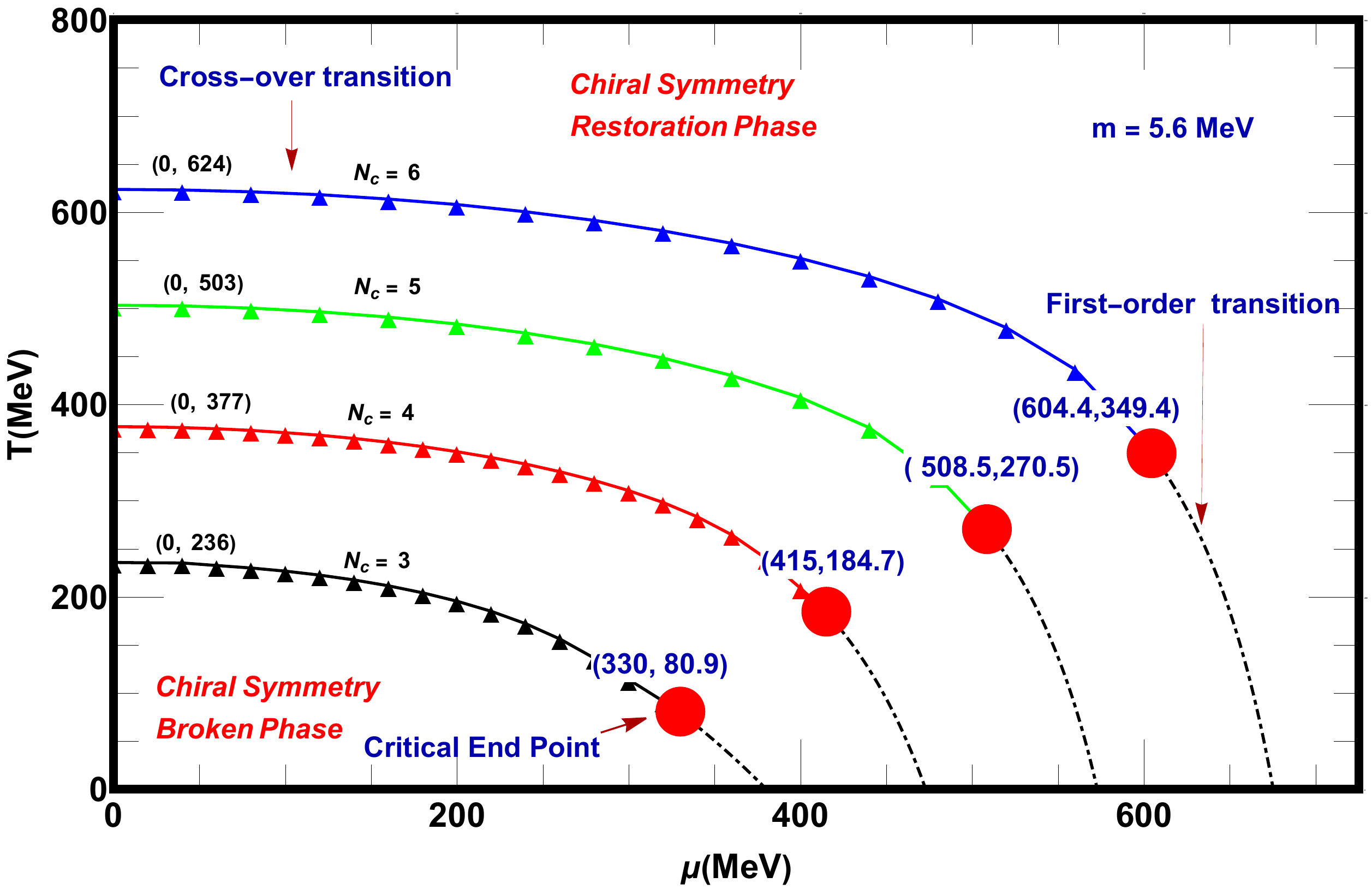}
\caption{QCD phase diagram for $T_c $ versus $\mu_c$, for various number of colors  $(N_c =3,4, 5,6)$  with fixed $ N_f = 2$. All solid-triangle line represent the cross-over phase transitions  and dot-dashed lines  for first order phase transition. The big red-dots for the critical endpoint in each phase diagrams. \label{QCDPNC}}
\end{center}
\end{figure}

\begin{figure}
\begin{center}
\includegraphics[scale=0.4]{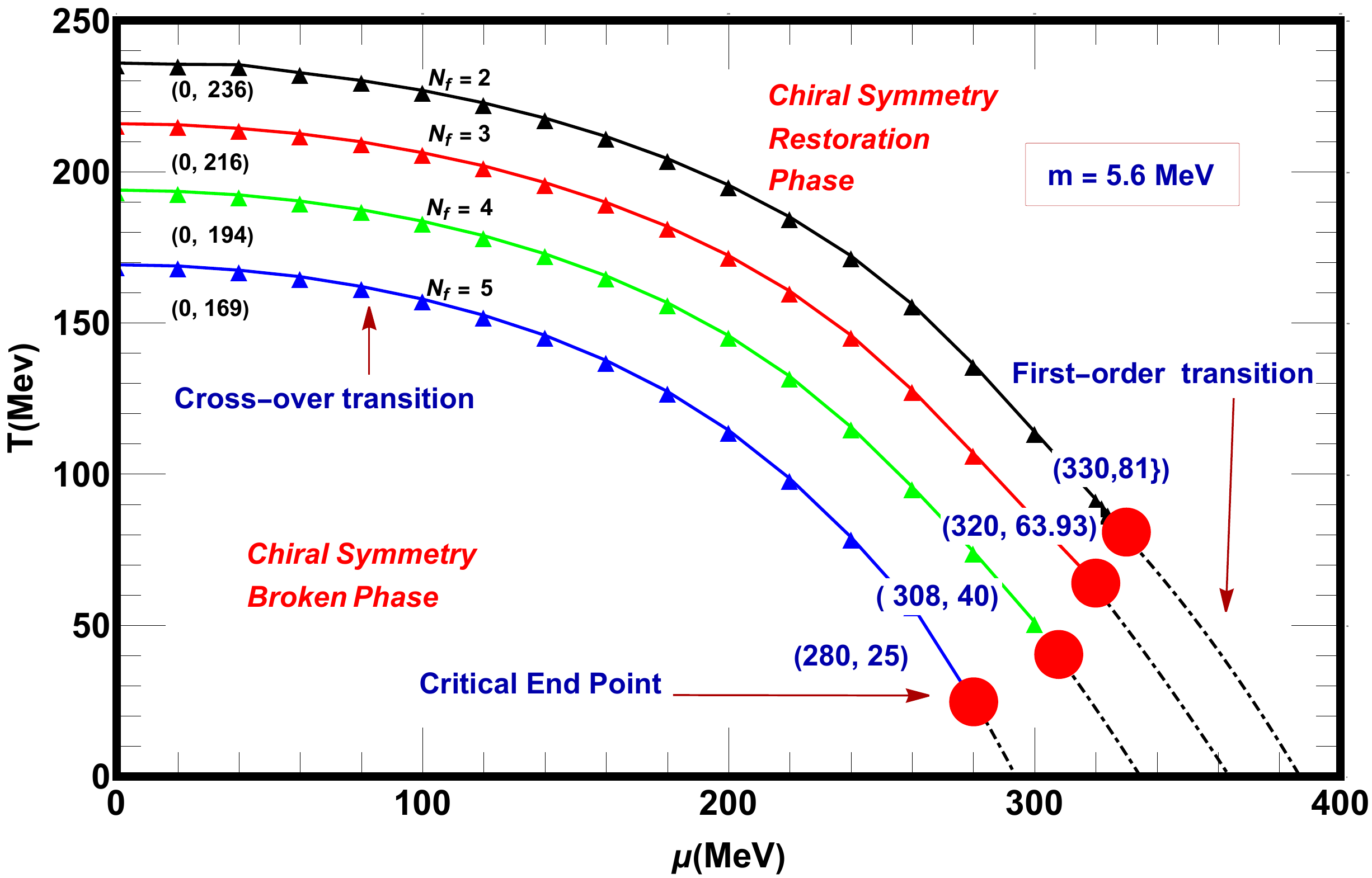}
\caption{QCD phase diagram for $T_c $ versus $\mu_c$, for various number of flavors  $(N_c  = 2,3,4,5)$ with fixed  number of colors $ N_c = 3$. All solid-triangle lines represent the cross-over phase transitions  and dot-dashed lines  for first order phase transition. The big red-dots for the critical endpoint in each phase diagram. \label{QCDPNF}}
\end{center}
\end{figure}

In this section, we sketched the QCD phase diagrams in the $T-\mu$ plane for various $N_c$ and $N_f$. First, we sketch the QCD phase diagram in the  Fig.~\ref{QCDPNC}, for fixed $N=2$ and for various number of colors (i.e, $N_c = 3, 4, 5, 6$), at finite temperature $T$ and chemical potential $\mu$. We obtained the critical temperature $T_c$ and critical chemical potential $\mu_c$ from the inflection points of the thermal and chemical potential gradients of quark-antiquark condensate, receptively. Initially, we plotted the phase diagram for $N_f = 2$ and $N_c = 3$, which  show that at finite $T$ but at $\mu= 0$, the dynamical chiral symmetry  broken for temperature $T\leq T_c \approx 235$ MeV, while above, it
is partially restored. The nature of the phase transition is
cross-over in this case. At finite $\mu$ and at $T = 0$, the dynamical symmetry is observed to be broken below $\mu_c \approx 380$ MeV, while above, it is
restored via first-order phase transition. It is confirmed from the Fig.~\ref{QCDPNC} that the cross-over line in the phase diagram started from
finite $T$-axis (for $\mu=0$) never ends up at finite $\mu$-axis (for $T=0$) and hence, there is a critical endpoint where the cross-over transition changes to first order. We determined the co-ordinates of the critical endpoint at
$(\mu^{E}_{c}\approx 330, T^{E}_{c}\approx 81)$ MeV. Our observation for $N_f = 2$ and $N_c = 3$ with a particular choice of  NJL model parameters,  
is consistent  with the
QCD phase diagram sketched in~\cite{buballa2005njl}. Then, we extended it 
to various higher number of colors $N_c$. The solid-triangular
line represents the cross-over phase transition for various $N_c$ phase diagrams and the dotted dashed line for
first order phase transition separated by the big red
dots (the critical endpoints) in all the phase diagrams. We 
observed that the $T_c$, $\mu_c$ and co-ordinates of the critical endpoints ($\mu^{E}_c $, $T^{E}_c$) are
shifted toward their higher values upon increasing $N_c$. This is because, the $N_c$ anti-screens the
interaction, while $T$ and  $\mu$  produced the screening effect. 
The variation of the critical temperature $T_c$ , the critical chemical potential $\mu_c$, and  the co-ordinates of critical endpoints ($\mu^{E}_c $, $T^{E}_c$) with the number of colors $N_c$ are
tabulated in the Tab.~\ref{QCDPNCt}.
\begin{table}[h!]
\begin{center}
\caption{Data for the  variation of the critical temperatures $ T_c $, the critical end point ($\mu^{E}_{c}, T^{E}_{c}$)in the phase diagrams with various  number of colors $ N_c $ and for fixed $N_f = 2$.  \label{QCDPNCt}.}
\begin{tabular}{|c|c|c|c|c|c|}
\hline
\textbf{S.No}&\textbf{$ N_c $}&\textbf{$m$}(MeV)&\textbf{$T_c$}(in MeV) at $\mu=0$&
\textbf{$ (\mu^{E}_{c}, T^{E}_{c})$}(MeV)\\[2pt]
\hline
 01  & 3  &  5.6  &  235 & (330, 81) \\[2pt]
 02  & 4  &  5.6  &  377 & (415, 185) \\[2pt]

 03  & 5  &  5.6  &  503 & (509, 271) \\[2pt]

 04  & 6  &  5.6  &  623 & (604, 349) \\[2pt]
\hline
\end{tabular}
\end{center}
\end{table}
Next, we draw the QCD
phase diagrams in the
Fig.~\ref{QCDPNF}, for fixed $N_c = 3$, but for various number
of light quark 
flavors (i.e., $N_f = 2,3, 4, 5$).
We used the same technique and same parameters for drawing the phase diagram, but now  we fixed
$N_c = 3$ and for various
number of light quark 
flavors $N_f$. All the phase diagrams in Fig.~\ref{QCDPNF}, shows their regular behaviors but suppresses with increasing number of flavors $N_f$
. We noticed that, upon varying 
flavors $N_f$, the critical temperature $T_c$, the critical chemical $\mu_c$ potential and the coordinates of the critical
endpoints ($\mu^{E}_c $, $T^{E}_c$) decreases with the increase of $N_f$
. This is because, all the three parameters $N_f$, $T$  and $\mu$ screens the interactions.  We show the
variation of the critical temperature $T_c$ , the critical chemical potential $\mu_c$, and  co-ordinates of critical endpoint ($\mu^{E}_c $, $T^{E}_c$) with the number of flavors  $N_f$ in the
the Tab.~\ref{QCDPNFt}. 
\begin{table}[h!]
\begin{center}
\caption{Data for the  variation of the critical temperatures $ T_c $, the critical end points ($\mu^{E}_{c}, T^{E}_{c}$) in the phase diagrams with various  number of  flavors $ N_f=2 $ and for fixed $N_c = 3$. \label{QCDPNFt}}
\begin{tabular}{|c|c|c|c|c|}
\hline
\textbf{S.No}&\textbf{$ N_f $}&\textbf{m}(MeV)&\textbf{$ T_c$ }(MeV) at $\mu=0$& CEP\textbf{$(\mu^{E}_{c}, T^{E}_{c})$} (MeV)\\[2pt]
\hline
 01   & 2 &  5.6 & 235&(330, 81) \\[2pt]
 02   & 3 &  5.6 & 216&(320, 64) \\[2pt]
 03   & 4 &  5.6 & 194&(308, 40) \\[2pt]
 04  & 5 &  5.6 & 169 &(280, 25) \\[2pt]
\hline
\end{tabular}
\end{center}
\end{table}\\
In the next section, we summarize our findings and draw the conclusions. 
\section{Summery and Conclusions} \label{section-VII} 
In this work, we have studied the dynamical chiral symmetry breaking/restoration for light quark flavors $N_f$ and 
colors $N_c$. Also, we investigated the impact of $N_c$ and  $N_f$ on the QCD phase diagram at finite temperature $T$ and quark chemical potential $\mu$.  For this purpose, we used the NJL model dressed with the color-flavor dependence of effective coupling $\mathcal{G}^{N_c}(N_f)$, which has advantages to study the QCD gap equation not only for $N_f=0$ but also for the higher number of flavors $N_f$. 
Our observations show that  for fixed $N_c=3$ and upon increasing $N_f$, the dynamical chiral symmetry is partially restored when $N_f$ exceeds a critical value $N_{f}^{c}\approx 8$. Our results have a remarkable resemblance with the modern Lattice QCD simulation and Schwinger-Dyson's equations predictions.  For $N_f=2$, upon increasing  the colors $N_c$, we determined the critical number of colors  $N_{c}^{c}\approx 2.2$, above which the dynamical chiral symmetry is broken. The dramatic opposed effects between the two parameters $N_{f}^{c}$ and $N_{c}^{c}$ has been observed. This is by our expectation and conformation of the previous studies \cite{Ahmad:2020jzn}, that is, increasing number of flavors $N_f$ screens the interactions while an increasing number of colors $N_c$ anti-screens them.\\  
At finite temperature $T$, our results show that the dynamical chiral symmetry is partially restored when $T$ reaches its critical value $T_c$. 
The temperature $T$ itself produces the screening effect in contrast to the anti-screening effect of colors $N_c$, consequently, the larger number of colors $N_c$ is required for the chiral symmetry breaking. Thus, the critical value $N^{c}_c$ (for $N_f = 2$) enhances as  $T$ increases.
For $N_c=3$, and upon increasing $N_f$, we find that both $T$ and  $N_f$, screens the interactions and hence, less number of
critical flavors $N^{c}_f$ needed to restored the  dynamical chiral symmetry. As a result, the  $N_{c}^{f}$ suppresses as $T$ increases. The nature of the  phase transition remains cross-over throughout for both ($N^{c}_c$ versus $T_c$) and  ($N^{f}_c$ versus  $T_c$) phase diagrams, respectively.
\\ Similarly, in the presence of quark chemical potential $\mu$, we concluded that the critical value of color $N_{c}^{c}$ for the dynamical chiral symmetry restoration, enhances as $\mu$ increases and vice verse. This is because the chemical potential $\mu$ itself screens the strong interaction. The nature of the phase transition, in this case, is observed to be smooth cross-over until the critical endpoint $(N^{cp}_{c}\approx2.5, \mu^{cp}_c\approx 290$ MeV) while above, the transition changes to the first order.   We further noticed that the critical number of light quark flavors $N^{c}_f$, required for the dynamical symmetry restoration suppressed with the increase of $\mu$,  because both  $N_f$ and $\mu$  produced the screening effect.  In this case, the nature of the phase transition is smooth cross-over until the critical endpoint $(N^{cp}_{f}\approx 5,\mu^{cp}_{f}\approx290$ MeV) while above, the transition changes to the first-order.\\
Finally, we sketched the QCD phase diagram at finite temperature $T-\mu$ plane, for various $N_c$ and $N_f$. We find that the critical temperature $T_c$, the critical chemical potential $\mu_c$ and the location of critical endpoint ($\mu^{E}_c $, $T^{E}_c$) shifted toward higher values, with the increasing number of colors $N_c$. While in case of increasing flavors $N_f$, the situation is opposite i.e., the critical temperature $T_c$,  the critical chemical potential $\mu_c$ and the critical endpoint ($\mu^{E}_c $, $T^{E}_c$) shifted towards their lower values with the increasing number of flavors $N_f$. We conclude that considering the number of light quark flavors (or colors) yields an important impact on the QCD phase diagram, besides the heat bath and background fields.  This work not only connects the color-flavor dependence with temperature $T$ and chemical potential  $\mu $ in congruence with the existing theoretical and phenomenological interpretation but also has important consequences related to the heavy-ion collision experiments.    
Soon, we plan to investigate the color-flavor phase diagram in the presence of background fields and other  light hadrons properties.
\section*{Acknowledgments}
We thanks to A. Bashir and A. Raya for their valuable
suggestions and guidance during the completion of this
manuscript. We acknowledge the organizers and participants of the ``$19th$ International Conference on Hadron Spectroscopy and Structure in memoriam Simon Eidelman (HADRON-$2021$), Mexico (online)'' and  ``International Union of on Pure and Applied Physics (IUPAP-$2022$) Pakistan (Online)'', for providing a nice environment for exchange of ideas during presentation of this work, which led to the
genesis of this manuscript. We also thanks to the colleagues of Institute of Physics, Gomal University for their
encouragement.
%\section{References}
\section*{References}
\bibliographystyle{iopart-num}
%\bibliographystyle{unsrt}
%\bibliography{aftab2}
%\bibliographystyle{apsrev4-1}
%\bibliographystyle{apacite}
\bibliography{aftab2}

\end{document}